\documentclass[secnumarabic,aps,prd,floats,floatfix,nofootinbib,showkeys,showpacs, twocolumn, 10pt]{revtex4-1}

\newcommand{\hs}{\hspace{0.01mm}}

\newcommand{\F}{\mathcal{F}}
\newcommand{\K}{\mathcal{K}}
\newcommand{\Planck}{{\it Planck} }

\usepackage{amsfonts}
\usepackage{amsmath}
\usepackage{amssymb}
\usepackage{natbib}
\usepackage{graphicx}
\usepackage{enumitem}
\usepackage[colorlinks]{hyperref}

\def\aj{AJ}
\def\apj{ApJ}
\def\apjl{ApJL}

\def\jcap{JCAP}
\def\mnras{MNRAS}
\def\physrep{Physics Reports}
\def\prd{Phys. Rev. D}
\def\apjs{ApJS}
\def\aap{A\&A }

\graphicspath{{./fig/}}

\begin{document}
	
\title{Cosmologically viable generalized Einstein-Aether theories}
\author{Richard~A.~\surname{Battye}}
\email{richard.battye@manchester.ac.uk}
\author{Boris~\surname{Bolliet}}
\email{boris.bolliet@manchester.ac.uk}
\author{Francesco~\surname{Pace}}
\email{francesco.pace@manchester.ac.uk}
\author{Damien~\surname{Trinh}}
\email[Corresponding author: ]{damien.trinh@postgrad.manchester.ac.uk}

\affiliation{Jodrell Bank Centre for Astrophysics, School of Physics and Astronomy, University of Manchester, 
	Manchester, M13 9PL, U.K.}

\label{firstpage}

\date{\today}

\begin{abstract}
We investigate generalized Einstein-Aether theories that are compatible with the \Planck Cosmic Microwave
Background (CMB) temperature anisotropy, polarisation, and lensing data. For a given dark energy equation of state, $w_{\rm de}$, we formulate a designer approach and we investigate their impact on the CMB temperature anisotropy and matter power spectra. We use the Equation of State approach to parametrize the perturbations and find that this approach is particularly useful in identifying the most suitable and numerically efficient parameters to explore in a Markov chain Monte Carlo (MCMC) \mbox{analysis}. We find the data constrains models with $w_{\rm de} = -1$ to be compatible with $\Lambda$CDM. For $w_{\rm de} \not= -1$ models, which avoid the gravitational waves constraint through the entropy perturbation, we constrain $w_{\rm de}$ to be $w_{\rm de } = -1.06^{+0.08}_{-0.03}$ (CMB) and $w_{\rm de } =-1.04^{+0.05}_{-0.02}$ (CMB+Lensing) at $68\%$C.L., and find that these models can be different from $\Lambda$CDM and still be compatible with the data. We also find that these models can ameliorate some anomalies in $\Lambda$CDM when confronted with data, such as the low-$\ell$ and high-$k$ power in the CMB temperature anisotropy and matter power spectra respectively, but not simultaneously. We also investigate the anomalous lensing amplitude, quantified by $A_{\rm lens}$, and find that that for $w_{\rm de} = -1$ models, $A_{\rm lens} = 1.15^{+0.07}_{-0.08}$ (CMB) and $A_{\rm lens} = 1.12\pm0.05$ (CMB+Lensing) at $68\%$C.L. $\sim$ 2$\sigma$ larger than expected, similar to previous analyses of $\Lambda$CDM.
\end{abstract}
\setlength{\abovedisplayskip}{12pt plus 2pt minus 9pt}
\setlength{\belowdisplayskip}{12pt plus 2pt minus 9pt}
%\pacs{98.80.-k, 95.36.+x}

%\keywords{Cosmology; vector field; Einstein-Aether; dark energy; equation of state}

\maketitle

\section{Introduction}

Cosmological observations suggest that we live in a Universe undergoing accelerated expansion, see for example \cite{Reiss1998SN, Perlmutter1999SN,Riess2007HST,Planck2015XIII}. Moreover, the data is consistent with a cosmological constant, $\Lambda$, as the origin for this acceleration \cite{Planck2016XIV}. Initial observations of type Ia supernovae allowed significant freedom for alternative dark energy and modified gravity models to explain the accelerated expansion \cite{Clifton2012MG,Ishak2018GR}. However, recent observations, and in particular of the propagation of gravitational waves \cite{LIGO2017GW1,LIGO2017GW2,LIGO2017GW3}, have greatly restricted the space of viable models. 

A popular set of scalar-tensor (ST) models are the Horndeski models \cite{Kobayashi2011Horn,Gleyzes2014Horn}, which are the most general that can be constructed up to second order derivatives in the scalar field. Its generality allows the testing of many subclasses of models such as Quintessence \cite{Peebles1988Q,Ratra1988Q,Caldwell1998Q}, k-essence \cite{Chiba2000KQ,Armend1999KQ}, $f(R)$ \cite{Sotiriou2006fR,Faulkner2007fR,Bean2007fR}, Kinetic Gravity Braiding (KGB) \cite{Deffayet2010KGB}, Galileons \cite{Deffayet2011Galileon}, and many others. Until very recently, the space of Horndeski models was relatively unconstrained, in that many subclasses yielded acceptable expansion histories and also were compatible with data from the CMB, large scale structure, and clustering. While cosmological data from these have helped in restricting some specific manifestations of these models, e.g. Galileons \cite{Barreira2013Galileon}, observations of gravitational waves, and more notably GW170817 and its electromagnetic counterpart GRB170817A \cite{LIGO2017GW1,LIGO2017GW2,LIGO2017GW3}, have been used to exclude many subclasses of Horndeski. In particular, the observations have constrained their propagation speed to be the speed of light, to very high precision. As several authors have pointed out, see for example \cite{Ezquiaga2017GW,Creminelli2017GW,Sakstein2017GW,Baker2017GW,Amendola2017GW}, this has left a significantly reduced model space of viable subclasses in Horndeski models and beyond. As it stands, current data is consistent with the $\Lambda$ Cold Dark Matter ($\Lambda$CDM) model and other alternatives are not significantly favoured.

That is not to say that the cosmological constant is itself without problems. It is well known that problems arise when interpreting $\Lambda$ as a vacuum energy in quantum field theory, often dubbed the naturalness problem \cite{Weinberg1989CC}. A related issue is the coincidence problem which is often discussed in arguments against a cosmological constant. This is related to why dark energy is only beginning to dominate today, despite its energy density having a very different evolution to that of matter and radiation, see for example \cite{Carroll2001CC} for a discussion. While alternative models do not necessarily themselves even solve these problems, at the very least it suggests our understanding of dark energy, whether its origin is the cosmological constant or not, is incomplete.

There are also a number of anomalies which currently exist within $\Lambda$CDM when confronted with data. Perhaps the most notable is the $\sim3\sigma$ discrepancy between the value of $H_0$ determined directly from local distance measures, $H_0 = (73.2\pm1.7)\, {\rm km \,s}^{-1}\, {\rm Mpc}^{-1}$ \cite{Riess2016H0}, and inferred from the angular scale of fluctuations in the CMB, $H_0 = (66.9 \pm 0.6)\, {\rm km \,s}^{-1}\, {\rm Mpc}^{-1}$ \cite{Planck2015XIII}. Another is that the data for the CMB temperature angular anisotropy power spectrum, $C_\ell^{\rm TT}$, for $\ell \lesssim 30$ is systematically below the prediction from $\Lambda$CDM \cite{Planck2016XX}, also similar to the data for the matter power spectrum, $P(k)$, at large $k$ \cite{Kitching2014CFHTLens,MacCrann2015s8}. Recent work has also highlighted the \Planck $A_{\rm lens}$ anomaly \cite{Calabrese2008Alens,Renzi2018Alens}, rescaling the lensing amplitude in the CMB spectra. This parameter is a consistency check and is not physically motivated, with an expected value of $1$. However, the latest \Planck analysis puts this value at $A_{\rm lens} = 1.15^{+0.13}_{-0.12}$ at 95$\%$C.L., $\sim2.3\sigma$ larger than expected \cite{Planck2016XLVI}. It is currently unknown whether these anomalies are due to systematics or new physics and provide more motivations for the field of modified gravity and dark energy.

While models like Horndeski introduce a dynamical scalar field instead of $\Lambda$ to modify General Relativity, an alternative is to consider the introduction of a vector field. Vector-tensor (VT) models of modified gravity have been shown to be capable of leading to periods of accelerated expansion and so provide an interesting line of dark energy research, complementary to those already studied in the context of the Horndeski class of models, see for example \cite{Beltran2009Vector,Jacobson2001EA,Zlosnik2007FK,Zuntz2010FK}. 

In this paper, we study the dynamics of cosmological perturbations in a class of VT models called generalized Einstein-Aether. First studied in \cite{Jacobson2001EA} and then generalized in \cite{Zlosnik2007FK}, these models introduce a vector field, $A^\mu$, known as the Aether field, that is constrained to have a timelike unit norm. Under this constraint, Einstein-Aether propagates only one scalar degree of freedom, similar to ST models. Its generalization comes about via non-canonical kinetic terms parametrized by a free function $\F(\K)$, where $\K$ is defined as
\begin{equation}
\mathcal{K} = \frac{1}{M^2}K^{\alpha \beta}\hs_{\mu \nu} \nabla_\alpha A^\mu \nabla_\beta A^\nu,
\end{equation} and the rank-4 tensor is given as
\begin{equation} \label{eq:Kinetic Tensor}
K^{\alpha \beta} \hs _{\mu \nu} = c_1 g^{\alpha \beta} g_{\mu \nu} + c_2\delta^\alpha _\mu \delta^\beta _\nu + c_3\delta^\alpha _\nu \delta^\beta _\mu + c_4 A^\alpha A^\beta g_{\mu \nu}.
\end{equation} Here, $M$ has dimensions of mass and $\left\lbrace c_i \right\rbrace$ are dimensionless constants, which are the free parameters of the theory. In particular, in this paper we study designer $\F(\K)$ models that mimic $\Lambda$CDM and $w$CDM background cosmologies but allow for the existence of non-trivial perturbations. These designer $\F(\K)$ models were studied in \cite{Battye2017GEA} for $w_{\rm de}=-1$. In considering such models, it is only the dynamics of the perturbations which will be important in distinguishing these models from $\Lambda$CDM. Of course, gravitational wave observations have also constrained this class of models and it can be shown that it places the restriction $c_1 + c_3 = 0$ or $d\F/d\K=0$ \cite{Baker2017GW}. The implications of this are discussed later on in the paper.

In recent years, a large amount of effort has been directed at developing parametrized frameworks for dark energy and modified gravity theories, in order to explore the theoretical landscape and departures from $\Lambda$CDM in a consistent manner. The philosophy behind this approach is to compress the freedom within the numerous different models of dark energy into a small number of phenomenological functions. These in turn can then be used to explore the parameter space of many different models. For example, in Horndeski  models a popular parametrization is via the Effective Field Theory approach and $\{\alpha_i\}$ functions, which can be related back to the physical properties of a given model \cite{Gubitosi2013EFT,Bellini2016alpha}. These approaches also include the Parametrized Post-Friedmann Framework \cite{Baker2013PPF,Skordis2015PPF}, a general theory of linear cosmological perturbations: ST and VT theories \cite{Lagos2016GT,Tattersall2017Cov} and the Equation of State (EoS) approach \cite{Battye2013EoS}, and the $(\mu,\gamma)$ or $(\mu, \Sigma)$ parametrization, for example see \cite{Zhao2010mg,Pogosian2016mg}. The main difference between these frameworks is the level at which they parametrize different models i.e. at the level of the action, the equations of motion, or the solutions to the equations of motion. Of course, when studying the effects of these models on cosmological observables, the choice of framework should not matter.

In this paper we work with the EoS approach, where the dark energy or modified gravity model is assumed to be a non-trivial cosmological fluid. At the level of linear perturbations, this approach eliminates the internal degrees of freedom introduced by the model and parametrizes the scalar sector via the gauge invariant anisotropic stress, $\Pi^{\rm S}$, and entropy perturbation, $\Gamma$, in order to close the perturbed conservation equations. Previous works have computed the equations of state for elastic dark energy \cite{Battye2007EDE}, $f(R)$ gravity \cite{Battye2016fR}, Quintessence, k-essence, KGB \cite{Battye2014EoS}, and more generally Horndeski theories \cite{Gleyzes2014Horn}. It was also applied to generalized Einstein-Aether theories in \cite{Battye2017GEA} and this paper continues that work by implementing this model in a modified version of \textsc{class} \cite{Blas2011CLASS}, called \textsc{class\textunderscore eos}, a modification to the Einstein-Boltzmann code that implements the EoS approach. For details of its numerical implementation we refer the reader to \cite{Battye2017FR}, where it was used to investigate designer $f(R)$ models. We will call the code used in this paper as \textsc{class\textunderscore eos\textunderscore gea}.

Previous works, such as \cite{Zuntz2008GEA, Zuntz2010GEA, Gong2018EA}, have attempted to constrain Einstein-Aether and generalized Einstein-Aether using observational data. Before the gravitational waves constraint, these models provided more compelling alternatives to $\Lambda$CDM. However, in light of recent constraints, these models have been severely restricted. In this work, we constrain generalized Einstein-Aether, in a similar way to \cite{Zuntz2008GEA, Zuntz2010GEA}, but instead of a power law solution for the general function $\F(\K)$, we opt for a designer approach. In particular, as well as studying models with $w_{\rm de} = -1$, which are now tightly constrained by gravitational waves, we investigate whether models with $w_{\rm de} \not= -1$ can still be observationally interesting, while still being compatible with the gravitational waves constraint. Such models will typically require $\Pi_{\rm de}^{\rm S} =0$ and so the modification to gravity is encoded solely by $\Gamma_{\rm de}$, though there are caveats to this, see for example \cite{Battye2018GW}. Since the current constraints from the data which apply to $\Gamma_{\rm de}$ are much weaker, we seek to investigate models whose modification to gravity comes about from a non-zero $\Gamma_{\rm de}$ only. The aim of this analysis is to understand how such models will affect cosmological observables and whether or not some of these models will be able to ameliorate some of the mentioned anomalies in $\Lambda$CDM. We will also investigate what the best parameters to explore are in a MCMC analysis, which we will see that the Equation of State approach is particularly useful for.

This paper is organized as follows. In \autoref{sect:Overview} we review generalized Einstein-Aether models and construct designer $\F(\K)$ models for a $w$CDM background. In \autoref{sect:EoS} we review the EoS approach to parametrizing the perturbations and apply this to generalized Einstein-Aether models. Using this approach, we study the evolution of dark energy perturbations in \autoref{sect:Evolve} and analyse their impact on cosmological observables in \autoref{sect:impact}. We then present observational constraints on the parameters in designer $\F({\K})$ models, from current CMB and lensing data in \autoref{sect:Spectra}. The effect of modifying the amplitude of the lensed $C_\ell$ via an $A_{\rm lens}$ parameter is also investigated. We then discuss our results and conclude in \autoref{sect:Conclusion}.

Natural units are used throughout with $c = \hbar = 1$ and the metric signature is $(-,+,+,+)$.

\section{Overview of generalized Einstein-Aether and designer $\F(\K)$} \label{sect:Overview}

In this section we briefly overview generalized Einstein-Aether theories and in particular, highlight important features of designer $\F(\K)$ models discussed in \cite{Battye2017GEA}. 

Generalized Einstein-Aether is defined by the action, in the Jordan frame,
\begin{equation} \label{eq:Action}
S = \int d^4 x \sqrt{-g} \left( \frac{1}{16 \pi G}R + \mathcal{L}_{\rm GEA}\right)  +S_m,
\end{equation} 
where
\begin{equation} \label{eq:Lagrangian}
16 \pi G\mathcal{L}_{\rm GEA} = M^2 \mathcal{F}(\mathcal{K}) + \lambda(g_{\mu \nu}A^\mu A^\nu +1).
\end{equation} 
The Lagrange multiplier term, $\lambda$, enforces the timelike unit norm constraint for the Aether field, $A^\mu$. Also note that $A^\mu$ does not couple directly to the matter sector. Variation of \eqref{eq:Action} with respect to the metric yields Einstein's equation in the form 
\begin{equation} \label{eq:EinsteinFieldEquation}
G_{\mu\nu} = 8\pi G T_{\mu\nu} + U_{\mu \nu},
\end{equation} 
where $T_{\mu \nu}$ is the standard matter energy-momentum tensor. Written in this way, all contributions from the Aether field are included in $U_{\mu\nu}$ and we will interpret this as the energy-momentum tensor of a non-trivial cosmological fluid. The full form of $U_{\mu\nu}$ is given in \cite{Battye2017GEA}. We assume a background cosmology described by the FLRW metric,
\begin{equation}
ds^2 = -dt^2 + a(t)^2\delta_{ij} dx^i dx^j,
\end{equation} 
and $A^\mu = (1,0,0,0)$ to be compatible with the symmetries from FLRW and also the timelike unit norm constraint. Projecting out the energy density, $\rho_{\rm GEA}$, and pressure, $P_{\rm GEA}$, we have that \begin{align} \label{eq:density}
&\rho_{\rm GEA} = 3 \alpha H^2 \left( \mathcal{F}_{\mathcal{K}} - \frac{\mathcal{F}}{2\mathcal{K}}\right), \\
\label{eq:Pressure}
P_{\rm GEA}= \alpha & \left[ 3H^2\left(\frac{\mathcal{F}}{2\mathcal{K}}-\mathcal{F}_{\mathcal{K}}  \right) - \dot{\mathcal{F}_{\mathcal{K}}} H -\mathcal{F}_{\mathcal{K}}\dot{H}\right],
\end{align} 
where overdots denote differentiation with respect to cosmic time, $t$, $H = \dot{a}/a$ is the Hubble factor, $\F_\K = d\F/d\K$, $\alpha = c_1 +3c_2 +c_3$, and for later use we will further define $c_{13} = c_1 +c_3$, $c_{14} = c_1 - c_4$, and $c_{123} = c_1 + c_2 + c_3$. We also have that 
\begin{equation} \label{eq:K}
\K = \frac{3\alpha H^2}{M^2}.
\end{equation} 
Note that due to the definition in \eqref{eq:EinsteinFieldEquation}, $\rho_{\rm GEA}$ and $P_{\rm GEA}$ have absorbed factors of $8\pi G$. We will therefore also define $8\pi G \rho_{\rm de} = \rho_{\rm GEA}$ and $8\pi G P_{\rm de} = P_{\rm GEA}$, where the  subscript `${\rm de}$' and, later on, `${\rm m}$' refers to dark energy and matter, respectively.

The freedom in the background evolution is currently governed by $\F(\K)$, its derivative, and $\{c_i\}$ via $\alpha$ and $\mathcal{K}$, as this dictates the evolution of $\rho_{\rm GEA}$, $P_{\rm GEA}$, and hence $w_{\rm de} = P_{\rm GEA}/\rho_{\rm GEA}$. One approach is to simply choose a form for $\F(\K)$ e.g. a power law as in \cite{Zuntz2010FK}, or more complicated functions as in \cite{Halle2008FK}. This would then allow us to fine-tune its functional form in order to be compatible with the observed background cosmology. Instead, we opt for a designer approach where we link the evolution of $a(t)$ with $\F(\K)$ so that $a(t)$ is identical to $\Lambda$CDM or $w$CDM. While this is somewhat artificial, it has the virtue that only the dynamics of the perturbations will be important in distinguishing these models from the standard $\Lambda$CDM and $w$CDM cosmologies. 

In \cite{Battye2017GEA}, it was found that for a given constant equation of state, $w_{\rm de}$, and energy density parameter today, $\Omega_{\rm de, 0} = 8\pi G\rho_{\rm de,0}/(3H_0^2)$, for a background cosmology indistinguishable from $w$CDM, $\F(\K)$ must obey
\begin{equation} \label{eq:BackgroundDifferentialEquation}
\begin{split}
(1&+w_{\mathrm{de}})\left( 2\mathcal{KF}_{\mathcal{K}}-\mathcal{F}\right) \\
 &= (2\mathcal{K}\mathcal{F}_\mathcal{KK}+\mathcal{F}_\mathcal{K})\left[ \mathcal{K} + \frac{1}{2}\alpha w_{\mathrm{de}} \left( 2\mathcal{K}\mathcal{F}_{\mathcal{K}}-\mathcal{F}\right) \right],
\end{split}
\end{equation} assuming negligible radiation contribution, which is true from matter domination and onwards, 
subject to the initial conditions 
\begin{equation} \label{eq:InitialConditions}
\quad \F(\K_0) = \F_0 \quad {\rm and} \quad \mathcal{F}_{\mathcal{K}}(K_0)= \frac{\Omega_{\rm{de},0}}{\alpha}+ \frac{\mathcal{F}_0}{2\mathcal{K}_0},
\end{equation} 
where $\K_0 = \K(a=1)$ and $\F_0$ is the value of $\F(\K)$ today, similar to the $B_0$ parameter in designer $f(R)$ theories. Solving \eqref{eq:BackgroundDifferentialEquation} will yield the behaviour for $\F(\K)$ required for a given $w_{\rm de}$ and $\Omega_{\rm de, 0}$, provided $\F_0$ is also given. It was shown in \cite{Battye2017GEA} that the background evolution of $\F(\K)$ was independent of the choice of $\{c_i\}$ and that $F_0$ was degenerate with $M$. The mass parameter, $M$, was therefore fixed at $M = H_0$ and will be for the rest of this paper.

In fact, \eqref{eq:BackgroundDifferentialEquation} is more complicated than what is required by \textsc{class\textunderscore eos\textunderscore gea}, which is $\F$ as a function of time, or some equivalent time variable e.g. scale factor or conformal time. Therefore, \eqref{eq:BackgroundDifferentialEquation} can be reduced to a first order equation in $a$ via the Friedmann equation, 
\begin{equation} \label{eq:ModifiedFriedmann}
(1- \alpha \mathcal{F}_{\mathcal{K}})\left( \frac{H}{H_0}\right) ^2 +\frac{1}{6}\mathcal{F}= \frac{\Omega_{\rm m, 0}}{a^3},
\end{equation} 
by demanding that the modifications to the Friedmann equation in \eqref{eq:ModifiedFriedmann} evolve as a general dark energy fluid with constant $w_{\rm de}$ i.e.
\begin{equation}
\alpha \mathcal{F}_{\mathcal{K}}\left( \frac{H}{H_0}\right) ^2 +\frac{1}{6}\mathcal{F}= \frac{\Omega_{\rm de, 0}}{a^{3(1+w_{\rm de})}}.
\end{equation} From \eqref{eq:K}, we can rewrite this as
\begin{equation} \label{eq:1ODE}
\F' = -\epsilon_H\left( \F +\frac{6 \Omega_{\rm de, 0}}{a^{3(1+w_{\rm de})}} \right),
\end{equation} 
where primes denote differentiation with respect to $\log a$ and $\epsilon_H = -H'/H$.

In \cite{Battye2017GEA}, it was found that the only cosmologically interesting solution to \eqref{eq:BackgroundDifferentialEquation} for $w_{\rm de} = -1$ was
\begin{equation} \label{eq:AnalyticalFminus1}
\mathcal{F}(a)= \left( \mathcal{F}_0 + 6\Omega_{\rm{de},0}\right)  \frac{H}{H_0}-6\Omega_{\rm{de},0},
\end{equation} 
which is consistent with \eqref{eq:1ODE}. Note that if $\F_0 = -6\Omega_{\rm{de},0}$ then $\F$ reduces to a constant and so this theory would be indistinguishable to $\Lambda$CDM at the perturbative level as well. For constant $w_{\rm de} \not= -1$, then the solution to \eqref{eq:1ODE} is given by 
\begin{equation} \label{eq:AnalyticalFnotminus1}
\frac{\F(a)}{6\Omega_{\rm de, 0}}=\frac{H}{H_0}\left( 1+ \frac{\F_0}{6\Omega_{\rm de, 0}}-\beta_0 \right)+a^{-3(1+w_{\rm de})}(\beta -1), 
\end{equation} 
where again we have assumed negligible radiation contribution to the total matter density. We have also further defined \begin{align}
\beta = \,&2\sqrt{1+a^{3w_{\rm de}}\frac{\Omega_{\rm m,0}}{\Omega_{\rm de, 0}}}\nonumber\\&\times{}_2F_1\left( \frac{1}{2},-\frac{1+w_{\rm de}}{2w_{\rm de}}; \frac{w_{\rm de}-1}{2w_{\rm de}}; -a^{3w_{\rm de}}\frac{\Omega_{\rm m, 0}}{\Omega_{\rm de, 0}} \right),
\end{align}
with $\beta_0 = \beta(a=1)$, and where ${}_2F_1(a,b;c;x)$ is the standard Gaussian hypergeometric function.
With the inclusion of radiation and ultra-relativistic species the solution is no longer analytical. However, given that we will start the dark energy perturbations well into the matter domination era, see \autoref{sect:Evolve}, this assumption is reasonable.

\section{Equation of State approach} \label{sect:EoS}

\begin{table}[ht]
	\caption{The dimensionless variables we choose to work with in the EoS approach are given in this table, in both the conformal Newtonian and synchronous gauges.} \label{table:Gauge Invariant Variables}
	\centering
	\begin{ruledtabular}
		\begin{tabular}{ c  c  c }
			Variable & Conformal Newtonian & Synchronous \\ \hline
			$T$ & $0$ & $\frac{h_{\parallel}'}{2\mathcal{H}{\rm K}^2}$\\
			$W$ & $\frac{1}{\mathcal{H}}X'-\epsilon_H(X+Y)$ & $\frac{1}{\mathcal{H}}X'-\epsilon_H(X+Y)$\\ 
			$X$ & $\frac{1}{\mathcal{H}}Z'+Y$ & $\frac{1}{\mathcal{H}}Z'+Y$  \\
			$Y$ & $\psi$ & $\frac{1}{\mathcal{H}}T'+\epsilon_H T$\\
			$Z$ & $\varphi$ & $\eta-T$\\ 
			$\Delta$ & $\delta + 3\mathcal{H}(1+w)\theta^{\rm S}$ & $\delta + 3\mathcal{H}(1+w)\theta^{\rm S}$ \\
			$\hat{\Theta}$&$3\mathcal{H}(1+w)\theta^{\rm S}$& $3\mathcal{H}(1+w)\theta^{\rm S} +3(1+w)T$ \\
			$\delta\hat{P}$& $\delta P$ & $\delta P +P'T$
		\end{tabular}
	\end{ruledtabular}
\end{table}

We will now briefly outline the Equation of State approach and its application to generalized Einstein-Aether models. Our starting point is \eqref{eq:EinsteinFieldEquation}, where we treat all modifications to General Relativity as a fluid via $U_{\mu \nu}$ and by construction, must be covariantly conserved i.e. $\nabla^\mu U_{\mu \nu} = 0$. 

At linear order in perturbations, $\delta U_{\mu \nu}$ is decomposed as 
\begin{equation} \label{eq:PerturbedEMT}
\begin{split}
\delta U^\mu\hspace{0.1mm}_{\nu} = \, &(\delta \rho + \delta P) u^\mu u_\nu + \delta P \delta^\mu \hspace{0.1mm}_{ \nu} \\ &+(\rho + P)(\delta u^\mu u_\nu + \delta u_\nu u^\mu) + P \Pi^\mu\hspace{0.1mm}_{\nu},
\end{split}
\end{equation} where the anisotropic stress, $\Pi^\mu\hspace{0.1mm}_{\nu}$, is further projected into scalar, vector, and tensor components via \cite{Ma1995Pert,Battye2006Pert}
\begin{equation} \label{Pi_ij}
\begin{split}
\Pi_{ij} = &\left( \hat{k}_i \hat{k}_j - \frac{1}{3}\delta_{ij} \right) \Pi ^{\rm S} + 2\hat{k}_{(i} \left( \Pi ^{\rm V1} \hat{l}_{j)} + \Pi^{\rm V2}\hat{m}_{j)}  \right) \\ &+ \Pi ^{+}\left( \hat{l}_i \hat{l}_j - \hat{m}_i \hat{m}_j   \right) +\Pi ^{\times}\left( \hat{l}_i \hat{m}_j -  \hat{l}_j  \hat{m}_i \right),
\end{split}
\end{equation}
where the unit vectors $\left\lbrace \hat{k}, \hat{l}, \hat{m} \right\rbrace $ form an orthonormal basis in $k$-space. 

The metric is perturbed as
\begin{equation}
ds^2 = a^2(\tau)\left[-(1+2\psi)d\tau^2+(\delta_{ij}+h_{ij})dx^idx^j \right],
\end{equation} in conformal time, $\tau$. Similar to $\Pi_{ij}$, $h_{ij}$ can be decomposed as in \eqref{Pi_ij}. This, together with the entropy perturbation,
\begin{equation}
w \Gamma = \left(\frac{\delta P}{\delta \rho} - \frac{dP}{d \rho} \right)\delta,
\end{equation} form the gauge invariant equations of state for the linear perturbations. In keeping with this gauge invariant language, we will work with a set of gauge invariant variables formed from the metric and perturbed fluid variables defined in \autoref{table:Gauge Invariant Variables}. Note that $T$  is not gauge invariant, but will not explicitly appear in the expressions for $\Pi^{\rm S}$ and $\Gamma$. For further details see \cite{Battye2016fR}. We also define the dimensionless wavenumber ${\rm K} = k/(aH)$.

After eliminating the internal degrees of freedom for the scalar sector, we write $\Pi^{\rm S}$ and $\Gamma$ as linear functions of the gauge invariant perturbed fluid variables, and find that
\begin{align} \label{eq:Pi2}
w_\mathrm{de}\Pi^{\rm S}_\mathrm{de} = \, & c_{\Pi\Delta}\Delta_\mathrm{de} + c_{\Pi\Theta}\hat{\Theta}_\mathrm{de}+c_{\Pi X}X \nonumber \\ &+ c_{\Pi Y}{\rm K}^2Y, \\
\label{eq:Gamma2}
w_\mathrm{de} \Gamma_{\mathrm{de}} = \, & c_{\Gamma\Delta}\Delta_\mathrm{de} + c_{\Gamma\Theta}\hat{\Theta}_\mathrm{de} +c_{\Gamma W}W \nonumber \\&+ c_{\Gamma X}X+ c_{\Gamma  Y}{\rm K}^2Y,
\end{align} where the $\{c_{\Pi,\Gamma}\}$ coefficients are in principle functions of both scale and time. 

The philosophy behind the EoS approach is that all modifications to gravity are treated as a new non-trivial cosmological fluid. To that end, we eliminate the metric variables $\{W,X,Y\}$ in favour of the perturbed fluid variables $\{\Delta_i, \hat{\Theta}_i\}$ via the Einstein equations, \begin{align} \label{eq:WEE}
2W = \, & \Omega_\mathrm{m} \left(\frac{3 \delta \hat{P}_\mathrm{m}}{\rho_\mathrm{m}} + 2 w_\mathrm{m}\Pi^{\rm S}_\mathrm{m} - 3 \hat{\Theta}_\mathrm{m} \right) \nonumber \\&+ \Omega_\mathrm{de} \left(\frac{3 \delta \hat{P}_\mathrm{de}}{\rho_\mathrm{de}} + 2 w_\mathrm{de}\Pi^{\rm S}_\mathrm{de} - 3 \hat{\Theta}_\mathrm{de} \right), \\ \label{eq:XEE}
2X = \, &\Omega_\mathrm{m} \hat{\Theta}_\mathrm{m} + \Omega_{\mathrm{de}} \hat{\Theta}_{\mathrm{de}}, \\ \label{eq:YEE}
-\frac{2}{3}{\rm K}^2 Y  = \, & \Omega_\mathrm{m} (\Delta_\mathrm{m} - 2w_\mathrm{m}\Pi^{\rm S}_\mathrm{m}) + \Omega_\mathrm{de} (\Delta_\mathrm{de} - 2w_\mathrm{de}\Pi^{\rm S}_\mathrm{de}), \\ \label{eq:ZEE}
-\frac{2}{3}{\rm K}^2 Z = \, & \Omega_\mathrm{m} \Delta_\mathrm{m} + \Omega_\mathrm{de} \Delta_\mathrm{de}.
\end{align} Substituting into \eqref{eq:Pi2} and \eqref{eq:Gamma2} yields
\begin{align} \label{eq:Pi Einstein}
w_\mathrm{de}\Pi^{\rm S}_\mathrm{de} = \,& c_{\Pi\Delta_\mathrm{de}}\Delta_\mathrm{de} + c_{\Pi\Theta_\mathrm{de}}\hat{\Theta}_\mathrm{de}+c_{\Pi\Delta_\mathrm{m}}\Delta_\mathrm{m} \nonumber \\&+ c_{\Pi\Theta_\mathrm{m}}\hat{\Theta}_\mathrm{m}+c_{\Pi\Pi_\mathrm{m}}\Pi_{\mathrm{m}}^{\rm S}, \\ \label{eq:Gamma Einstein}
w_\mathrm{de} \Gamma_{\mathrm{de}} = \,& c_{\Gamma\Delta_\mathrm{de}}\Delta_\mathrm{de} + c_{\Gamma\Theta_\mathrm{de}}\hat{\Theta}_\mathrm{de} +c_{\Gamma\Delta_\mathrm{m}}\Delta_\mathrm{m} \nonumber \\&+ c_{\Gamma\Theta_\mathrm{m}}\hat{\Theta}_\mathrm{m}+ c_{\Gamma \Gamma_{\mathrm{m}}}\Gamma_{\mathrm{m}},
\end{align} where we have included $\Pi_{\rm m}$ and $\Gamma_{\mathrm{m}}$ for generality, as these are non-zero for ultra-relativistic matter species. Note that $c_{\Pi\Delta}$ and $c_{\Pi\Delta_{\rm de}}$ are not the same since $Y$ can be rewritten in terms of $\left\lbrace \Delta_i\right\rbrace $ via \eqref{eq:YEE}, similarly for $c_{\Gamma\Delta}$ and $c_{\Gamma\Delta_{\rm de}}$. The relationship between these coefficients and the previous ones are given in Appendix \ref{app:A}.

Their forms in generalized Einstein-Aether were computed in \cite{Battye2017GEA} in full generality and are given in Appendix \ref{app:A}, however they simplify significantly for $w_{\rm de}=~-1$ in designer $\F(\K)$ models. From \eqref{eq:AnalyticalFminus1} we have an analytical solution and in this case the $\{c_{\Pi,\Gamma}\}$ coefficients reduce to 
\begin{equation} \label{eq:w-1coeffs}
\begin{split}
&c_{\Pi\Delta} = \frac{c_{13}}{c_{14}}, \quad
c_{\Pi\Theta} = \frac{1}{2}\left(1+\epsilon_H \right) - \frac{c_{13}}{c_{14}}, \\
&c_{\Pi X} = 0,  \, c_{\Pi Y} =  -\frac{c_{13}}{3\alpha}\left( 1+\frac{\mathcal{F}_0}{6\Omega_{\mathrm{de},0}}\right)\frac{H}{H_0},
\end{split} 
\end{equation} and \begin{equation}
c_{\Gamma \Delta} = -c_{\Gamma \Theta} = -\frac{dP_{\rm de}}{d\rho_{\rm de}}=1, \quad
c_{\Gamma W} = c_{\Gamma X} = c_{\Gamma Y} = 0.
\end{equation} Together, these coefficients completely encode the modification to gravity due to a designer $\F(\K)$ model with $w_{\rm de} = -1$.

In a designer background, $H$ is already determined and so the only two parameters which will dictate how these models impact cosmological observables are \begin{equation} \label{eq:P1 and P2}
\mathcal{P}_1= \frac{c_{13}}{c_{14}} \quad {\rm and} \quad \mathcal{P}_2 = \frac{c_{13}}{\alpha}\left( 1+\frac{\mathcal{F}_0}{6\Omega_{\mathrm{de},0}}\right),
\end{equation} and it will be these parameters that we will explore over in the MCMC analysis, see \autoref{sect:Spectra}. Note that we have reduced the initial 5 free parameters of this theory, i.e. $\left\lbrace c_i, \F_0\right\rbrace $, to only 2. In principle, we could run the MCMC analysis with these 5 parameters, but this would only show us that degeneracies exist between these parameters, as shown by  \eqref{eq:P1 and P2}, and so it is more numerically efficient to directly use the 2 parameters $\mathcal{P}_1$ and $\mathcal{P}_2$. This is one of the advantages of the EoS approach, in that it allows us to see explicitly what combinations of parameters will directly affect the observables and what degeneracies exist between the original parameters. Note that imposing flat priors on $\mathcal{P}_1$ and $\mathcal{P}_2$ is not equivalent to doing the same for the original 5 parameters.

For $\F_0 = -6\Omega_{\rm de, 0}$ we have that $\mathcal{P}_2 = 0$. This case is indistinguishable from $\Lambda$CDM at the level of linear perturbations as $\F$ is constant, corresponding to $\F_\K = 0$. As we will see later, another case that recovers $\Lambda$CDM is when $c_{13}=0$ and so $\mathcal{P}_1 = \mathcal{P}_2=0$, see \autoref{sect:Evolve}.

Note that in \cite{Battye2017GEA} it was found that
\begin{equation} \label{cs2}
c_{\rm s}^2 = \frac{1}{c_{14}}\left( c_{123} + \frac{2}{3}\alpha\gamma_2\right),
\end{equation} could be interpreted as a sound speed for perturbations i.e. the coefficient of $k^2 \delta_{\rm de}$ in the $\ddot{\delta}_{\rm de}$ equation, however it need not necessarily be so. After all, a sound speed is itself frame dependent. For $w_{\rm de} = -1$ models, we find that $c_{\rm s}^2 = \frac{2}{3}\mathcal{P}_1$. Indeed, \eqref{cs2} is also consistent with \cite{Lim2004EA,Jacobson2004EA} where they computed the wave speed of different modes in Einstein-Aether, in the Minkowski limit. However, as discussed in \cite{Battye2017GEA}, in designer $\F(\K)$ models where we have directly coupled the evolution of $\F$ to $a(t)$, no sensible Minkowski limit exists for this theory once this connection has been made. It could be argued that on grounds of subluminal propagation $P_1$ should have a upper bound of $\frac{3}{2}$. However, as previously mentioned it is not necessarily the sound speed and if it was it would only refer to the phase velocity. We will therefore leave the upper bound of $\mathcal{P}_1$ unrestricted.

In complete generality it is not clear what the equivalent parameters to $\mathcal{P}_1$ and $\mathcal{P}_2$ are when $w_{\rm de} \not= -1$. However, we will see in the next section that the gravitational wave constraints significantly reduce the complexity of such models, allowing us to identify the parameters in a general way.

\subsection{Constraints from gravitational waves}

Recent observations of gravitational waves, and in particular detections with a coincident gamma ray burst, have provided strong constraints on the deviation of the speed of gravitational waves, $c_{\rm grav}$, from light, $c_\gamma$. Specifically, GW170817 and GRB170817A \cite{LIGO2017GW1,LIGO2017GW2,LIGO2017GW3} have constrained this deviation to be \begin{equation}
-3\times 10^{-15} < \frac{c_{\rm grav}-c_\gamma}{c_\gamma}<7 \times 10^{-16}.
\end{equation} While there may be caveats to this constraint, see \cite{Battye2018GW}, taken as it is, it suggests that gravitational waves must propagate at the speed of light. In generalized Einstein-Aether, $c^2_{\rm grav}$ is given by
\begin{equation}
c^2_{\rm grav}=\left( 1 + c_{13}\F_\K \right)^{-1} ,
\end{equation} and hence we require $c_{13}=0$ or $\F_\K = 0$. The latter is the case of a cosmological constant and so we will focus on the potentially more interesting case of $c_{13}=0$.

From \eqref{c_PiD} - \eqref{c_PiY} we see that all $\left\lbrace c_{\Pi}\right\rbrace$ coefficients are proportional to $c_{13}$ and hence the gravitational waves constraint sets $\Pi^{\rm S}_{\rm de} = 0$. Therefore, the modification to gravity in these models is encoded solely in $\Gamma_{\mathrm{de}}$. Under this constraint, the $\left\lbrace c_{\Gamma} \right\rbrace$ coefficients \eqref{c_GammaD} - \eqref{c_GammaY} can be written as \begin{equation}\label{eq:wnot-1coeffs}
c_{\Gamma \Delta} = \frac{3(1+w_{\rm de})}{\epsilon_H\left[ a^{4+3w_{\rm de}}(\mathcal{P}_4- \mathcal{P}_3\beta_0)\left( \frac{H}{H_0}\right)+a\mathcal{P}_3 \beta\right] } - w_{\rm de},
\end{equation}
\begin{align}
c_{\Gamma \Theta} &= \frac{2}{3}(\epsilon_H-1)-\frac{\epsilon_H'}{3\epsilon_H}, \\
c_{\Gamma W} &= \frac{1}{2}c_{\Gamma X} = -3 c_{\Gamma Y} = \frac{1+w_{\rm de}}{\epsilon_H}, 
\end{align}  
and the new parameters we choose to explore are 
\begin{equation}
\mathcal{P}_3 = \frac{c_{14}}{c_{2}} \quad {\rm and} \quad
\mathcal{P}_4 = \frac{c_{14}}{c_{2}}\left( 1 + \frac{\F_0}{6\Omega_{\rm de, 0}}\right).
\end{equation} These choices are motivated by the forms of $\mathcal{P}_1$ and $\mathcal{P}_2$ in \eqref{eq:P1 and P2} for $w_{\rm de} = -1$ models.

In principle, the gravitational waves constraint of either $\F_\K$ or $c_{13} = 0$ could also be applied to $w_{\rm de} = -1$ models. However, as mentioned previously, this would correspond to $\Lambda$CDM at the level of perturbations, rendering the MCMC analysis for such models unnecessary. In such cases, either $\mathcal{P}_1 \not= 0$ but $\mathcal{P}_2 =0$, corresponding to $\F_\K = 0$, where such models are degenerate with $\Lambda$CDM for any value of $\mathcal{P}_1$, or $\mathcal{P}_1 = \mathcal{P}_2 =0$, corresponding to $c_{13}=0$, and so there are no parameters to explore. However, it will still be instructive to consider the dynamics of $w_{\rm de} = -1$ models without the gravitational waves constraint applied, which will aid our understanding of the dynamics when $w_{\rm de}\not=-1$ and see whether the MCMC analysis will pick out $w_{\rm de} = -1$ models that are consistent with either $\F_\K = 0$ or $c_{13}=0$, without information from gravitational waves.

\section{Cosmological dynamics} \label{sect:Evolve}
We now move on to investigate the evolution of cosmological perturbations, both analytically and numerically, in designer $\F(\K)$ models. 
\subsection{Dynamics of linear perturbations}
For simplicity we will assume that radiation is negligible, which is true for the times  that we are interested in and so $\Pi^{\rm S}_{\rm m} = \Gamma_{\rm m} = 0$. The perturbed conservation equation gives 2 coupled first order differential equations for each species, given by \begin{align} \label{eq:ConsEq1GI}
\Delta'-3w\Delta+g_{\mathrm{K}}\epsilon_H \hat{\Theta}-2w\Pi^{\rm S} &= 3(1+w) X, \\ \label{eq:ConsEq2GI}
\hat{\Theta}' + 3\left(\frac{dP}{d\rho} - w + \frac{1}{3} \epsilon_H \right)\hat{\Theta} \nonumber \\- 3\frac{dP}{d\rho}\Delta - 2w\Pi^{\rm S} - 3w\Gamma &=3(1+w)Y,
\end{align} where $g_{\mathrm{K}}=1+\frac{{\rm K}^2}{3\epsilon_H}$. The initial conditions are chosen such that dark energy perturbations are negligible at $z_{\rm ini}=100$ i.e. $\Delta_{\mathrm{de}} = \hat{\Theta}_\mathrm{de} = 0$, $\Omega_{\mathrm{m}}\Delta_{\mathrm{m}} = -\frac{2}{3} {\rm K}^2 Z$, $\Omega_{\mathrm{m}}\hat{\Theta}_\mathrm{m}=2X$, and $X=Y=Z$. The exact starting point for the evolution of the dark energy perturbations is somewhat arbitrary provided we are sufficiently into the matter dominated era. If this is the case then the results are not sensitive to the precise value of $z_{\rm ini}$.

\begin{figure}
	\centering
	\includegraphics[width=0.5\textwidth]{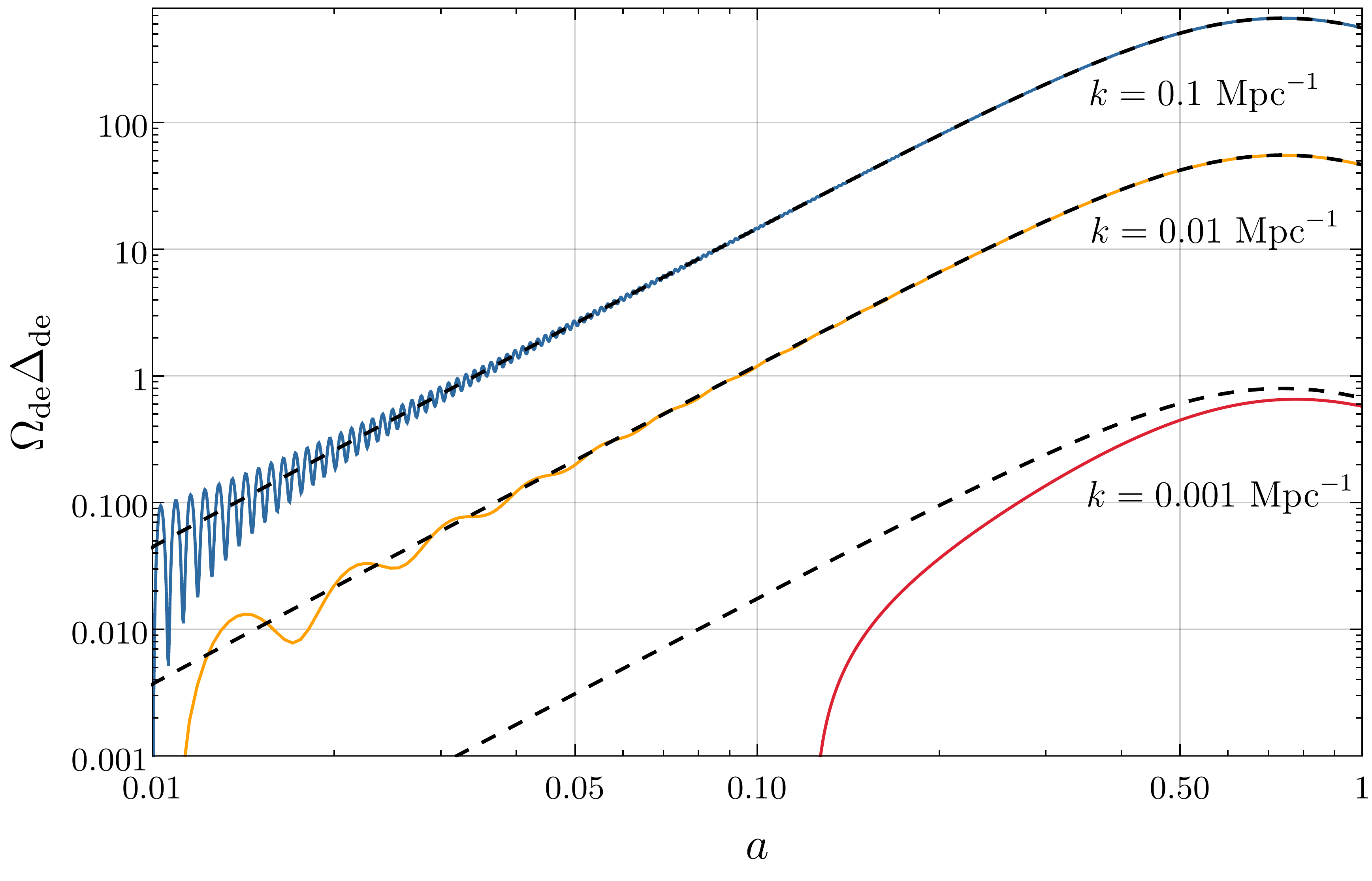}
	\caption{ The evolution of $\Omega_{\rm de}\Delta_{\rm de}$ is shown for $k = 0.001 \ {\rm Mpc}^{-1}$, $0.01 \ {\rm Mpc}^{-1}$, and $0.1 \ {\rm Mpc}^{-1}$, for $\log \mathcal{P}_1 = 1.0$ and $\mathcal{P}_2 = 1.0$ fixed, compared with the attractor solution for each scale (black dashed lines).}
	\label{fig:DeltaDE}
\end{figure}

 \begin{figure}
	\centering
	\includegraphics[width=0.5\textwidth]{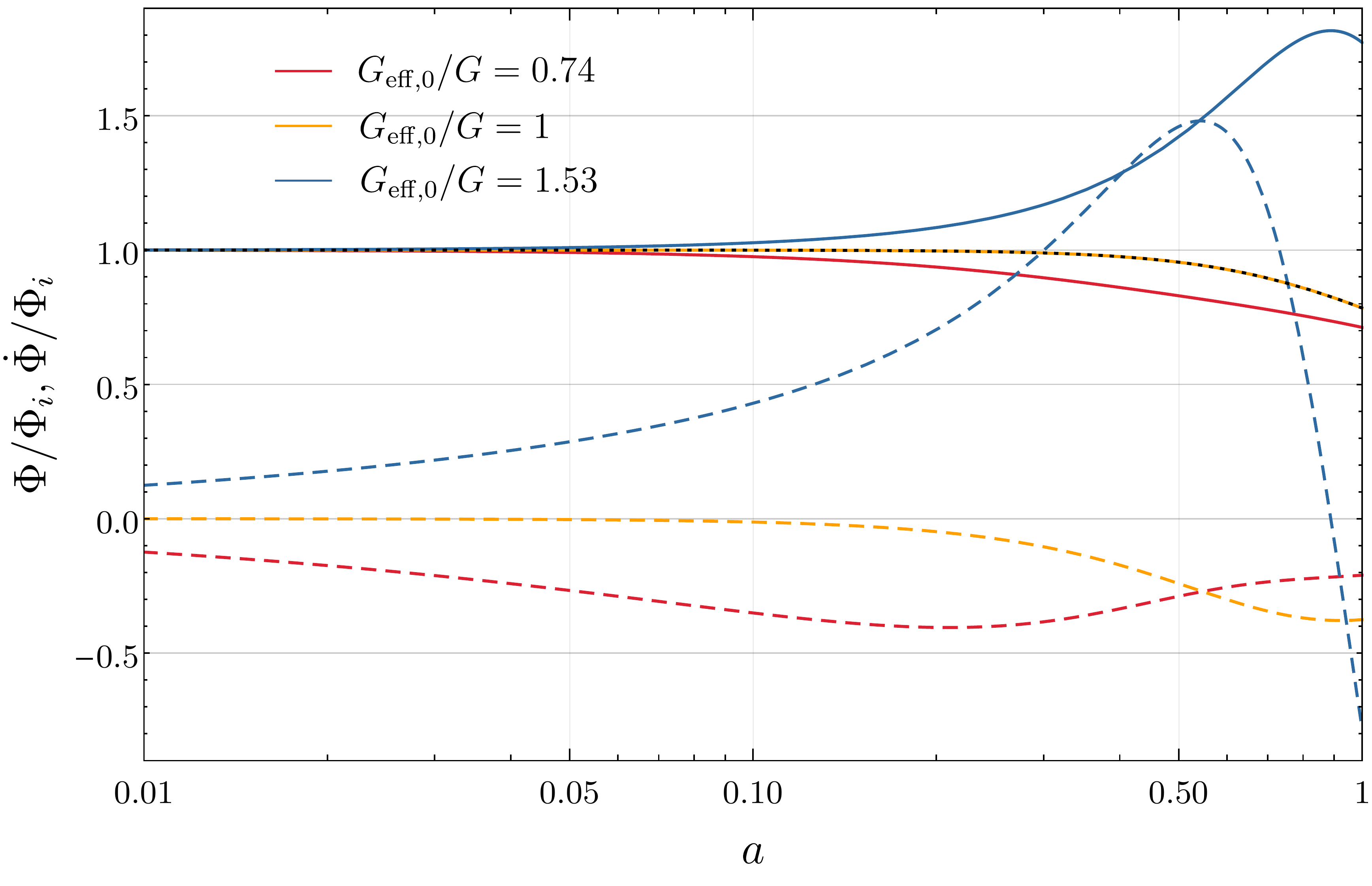}
	\caption{The Weyl potential, $2\Phi = \psi + \phi$, (solid lines) and its derivative (dashed lines) as a function of scale factor for different values of $G_{\rm eff,0}$, corresponding to $\mathcal{P}_2 = 1$, $0$, and $-1$, with $\log \mathcal{P}_1=0.1$, $w_{\rm de} = -1$, and ${\rm K}_0 = 1$ fixed. The  $\Lambda$CDM potential is denoted by the black dotted line.}
	\label{fig:Phi}
\end{figure}

The behaviour of the Newtonian potentials was studied in \cite{Battye2017GEA} and in particular it was found that the gravitational slip $\eta = \phi/\psi \rightarrow 1$ for ${\rm K} \gg 1$ i.e. $\phi = \psi$, or equivalently $\Pi^{\rm S}_{\rm de} = 0$, for ${\rm K} \gg 1$. Therefore, we would expect complete consistency with $\Lambda$CDM in the CMB temperature angular anisotropy power spectrum at high-$\ell$. However, for low-$\ell$, we would expect differences as the late-time Integrated Sachs-Wolfe (ISW) effect is sensitive to $\Pi^{\rm S}_{\rm de}$ and, in particular, to the variation of $\phi$ and $\psi$. Furthermore, it is these large scale modes that enter the horizon at late times when the dark sector component is dominating. It should be noted that if $\Pi^{\rm S}_{\rm de} \not =0$ for $K \gg 1$, then it is still possible to be consistent with $\Lambda$CDM at high-$\ell$, see, for example, \cite{Battye2017FR} in $f(R)$.

In \cite{Battye2017GEA} it was found that for ${\rm K} \gg 1$ the perturbed conservation equations, for $w_{\rm de}=-1$, could be written as 2 coupled second order differential equations, given by \begin{align} \label{eq:2ODE DeltaM}
\Delta''_\mathrm{m} +(2-\epsilon_H)\Delta'_\mathrm{m} - \frac{3}{2}\Omega_{\mathrm{m}}\Delta_{\mathrm{m}}&=\frac{3}{2}\Omega_{\mathrm{de}}\Delta_{\mathrm{de}},\\ \label{eq:2ODE DeltaDE}
\Delta''_\mathrm{de}+(5-\epsilon_H)\Delta'_{\mathrm{de}} +\frac{2}{3}c_{\Pi\Delta_{\mathrm{de}}}{\rm K}^2\Delta_{\mathrm{de}}&=-\frac{2}{3}c_{\Pi\Delta_{\mathrm{m}}}{\rm K}^2\Delta_{\mathrm{m}},
\end{align} provided the $\{\hat{\Theta}_i\}$ terms were small relative to the $\left\lbrace \Delta_i \right\rbrace $ terms. This will be true for small scales where ${\rm K} \gg 1$, as can be seen from the Einstein equations \eqref{eq:XEE} and \eqref{eq:YEE}. This regime corresponds to modes which are within the horizon during matter domination. From \eqref{eq:2ODE DeltaDE} we see that $\Delta_{\rm de}$ will tend to the attractor solution
\begin{equation} \label{eq:Attractor}
\Delta_{\mathrm{de}} = -\frac{c_{\Pi\Delta_{\mathrm{m}}}}{c_{\Pi\Delta_{\mathrm{de}}}}\Delta_{\mathrm{m}},
\end{equation} shown in \autoref{fig:DeltaDE}. The $\Lambda$CDM background cosmology was set such that $h=H_0/ 100 \, {\rm km \,s}^{-1}\, {\rm Mpc}^{-1} = 0.68$, $\Omega_{\rm de, 0} = 0.69$, and $\Omega_{\rm b}h^2 = 0.022$. We see that initially $\Delta_{\rm de}$ grows to match $-c_{\Pi\Delta_{\mathrm{m}}}/c_{\Pi\Delta_{\mathrm{de}}}\Delta_{\mathrm{m}}$ and oscillates about the attractor solution. These oscillations are a consequence of setting the dark energy perturbations to be zero at $z=100$. If the initial conditions are set at earlier times, the amplitude of the oscillations is suppressed since the attractor solution will be closer to zero. We will therefore set the initial condition for $\Delta_{\rm de}$ in \textsc{class\textunderscore eos\textunderscore gea} to match the attractor. This is numerically more efficient and in doing so the oscillations will be suppressed.

Substituting \eqref{eq:Attractor} into \eqref{eq:2ODE DeltaM} yields the standard evolution equation for $\Delta_{\rm m}$ in $\Lambda$CDM, but with an effective Newtonian gravitational constant, given by 
\begin{equation}
\frac{G_{\mathrm{eff}}}{G} = 1 - \frac{\Omega_{\mathrm{de}}c_{\Pi\Delta_{\mathrm{m}}}}{\Omega_{\mathrm{m}}c_{\Pi\Delta_{\mathrm{de}}}}.
\end{equation} This parameter is important in explaining the dynamics of cosmological perturbations and hence the cosmological observables. We can write the value of $G_{\rm eff}$ today as \begin{equation} \label{eq:Geff}
\frac{G_{\rm eff,0}}{G} = \frac{1}{1+\frac{\mathcal{P}_2}{2\mathcal{P}_1}\Omega_{\rm de, 0}}.
\end{equation} As mentioned previously, if $\mathcal{P}_2 = 0$, corresponding to $\F_\K=0$, then we recover $\Lambda$CDM and we find that $G_{\rm eff,0}=G$ and in fact so is $G_{\rm eff}(a)$ for all $a$. As we will see later, for $c_{13}=0$ where $\mathcal{P}_1 = \mathcal{P}_2 = 0$, there are no growing modes in $\Delta_{\rm de}$ and hence we recover the standard $\Lambda$CDM evolution equation in \eqref{eq:2ODE DeltaM} upon setting $\Delta_{\rm de} = 0$, with the standard Newtonian gravitational constant.

A similar analysis can applied to $w_{\rm de} \not=-1$ models. We find that, for ${\rm K}\gg1$, \eqref{eq:2ODE DeltaM} still holds, but now
\begin{align} \label{eq:2ODE wDeltaDE}
\Delta_{\rm de}''&+\left[ (2-3w_{\rm de})-\epsilon_H\right] \Delta_{\rm de}'\nonumber\\&+(w_{\rm de}+c_{\Gamma \Delta_{\rm de}}){\rm K}^2\Delta_{\rm de} = -c_{\Gamma\Delta_{\rm de}}{\rm K}^2 \Delta_{\rm m}.
\end{align} Hence, the attractor solution is modified to \begin{equation}
\Delta_{\rm de} = -\frac{c_{\Gamma\Delta_{\rm m}}}{w_{\rm de}+c_{\Gamma\Delta_{\rm de}}}\Delta_{\rm m},
\end{equation} and the effective Newtonian gravitational constant becomes
\begin{equation} 
\frac{G_{\rm eff}}{G}=1-\frac{\Omega_{\rm de}c_{\Gamma\Delta_{\rm m}}}{\Omega_{\rm m}(w_{\rm de}+c_{\Gamma\Delta_{\rm de}})}.
\end{equation} Note that the value of $G_{\rm eff,0}$ in these models can be written as
\begin{equation} \label{eq:Geff2}
\frac{G_{\rm eff,0}}{G} = \frac{6}{6+\mathcal{P}_4\Omega_{\rm de, 0}},
\end{equation} i.e. $G_{\rm eff,0}$ is independent of $w_{\rm de}$ and depends only on $\mathcal{P}_4$ and in principle $\mathcal{P}_3$, since 
\begin{equation} \label{eq:P3P4}
\mathcal{P}_4 = \left( 1+ \frac{\F_0}{6\Omega_{\rm de,0}}\right) \mathcal{P}_3.
\end{equation}

Note that there is not a direct link between $\left\lbrace \mathcal{P}_1, \mathcal{P}_2\right\rbrace $ and $\left\lbrace \mathcal{P}_3, \mathcal{P}_4\right\rbrace $, since the former parametrizes $\Pi^{\rm S}_{\rm de}$ in $w_{\rm de} = -1$ models where $\Gamma_{\rm de}$ is fixed and contains no free parameters and the latter parametrizes $\Gamma_{\rm de}$ in $w_{\rm de} \not =-1$ models where $c_{13}=0$ and so $\Pi^{\rm S}_{\rm de} =0$.

 \subsection{Dynamics under the gravitational wave constraint}
 
 Consider again the gravitational waves constraint which demand either $\F_\K=0$ or $c_{13}=0$. It is clear that the case of $\F_\K=0$ will yield a model identical to $\Lambda$CDM and hence there will be no dark energy perturbations. However, we find that this is also true if $c_{13} = 0$ and $w_{\rm de}= -1$. 
 
 To see this, recall that $\Pi_{\rm de}^{\rm S} = 0$ if $c_{13} = 0$ and so \eqref{eq:2ODE DeltaDE} becomes \begin{equation} \label{eq:2ODE_gw}
 \Delta''_\mathrm{de}+(5-\epsilon_H)\Delta'_{\mathrm{de}} = 0.
 \end{equation} In the matter dominated era $\epsilon_H \rightarrow 3/2$ and so the solution to \eqref{eq:2ODE_gw} has no growing modes, i.e. if $c_{13} = 0$ then there are no dark energy perturbations, since any non-zero initial condition for $\Delta_{\rm de}$ would quickly decay. Therefore, the constraint that $c_{13}=0$ restricts designer $\F(\K)$ models to those which are indistinguishable from $\Lambda$CDM both at the level of background cosmology and linear perturbations, if $w_{\rm de} = -1$. Therefore, to explore cosmologically interesting models which are different from $\Lambda$CDM and compatible with the gravitational waves constraint, we cannot restrict ourselves to $w_{\rm de} = -1$. However, as mentioned previously, it will still be interesting and instructive to study the impact of such models on cosmological observables which we can use to aid our understanding of models with $w_{\rm de} \not= -1$.
 
  \begin{figure*}
 	\centering
 	\includegraphics[width=0.495\textwidth]{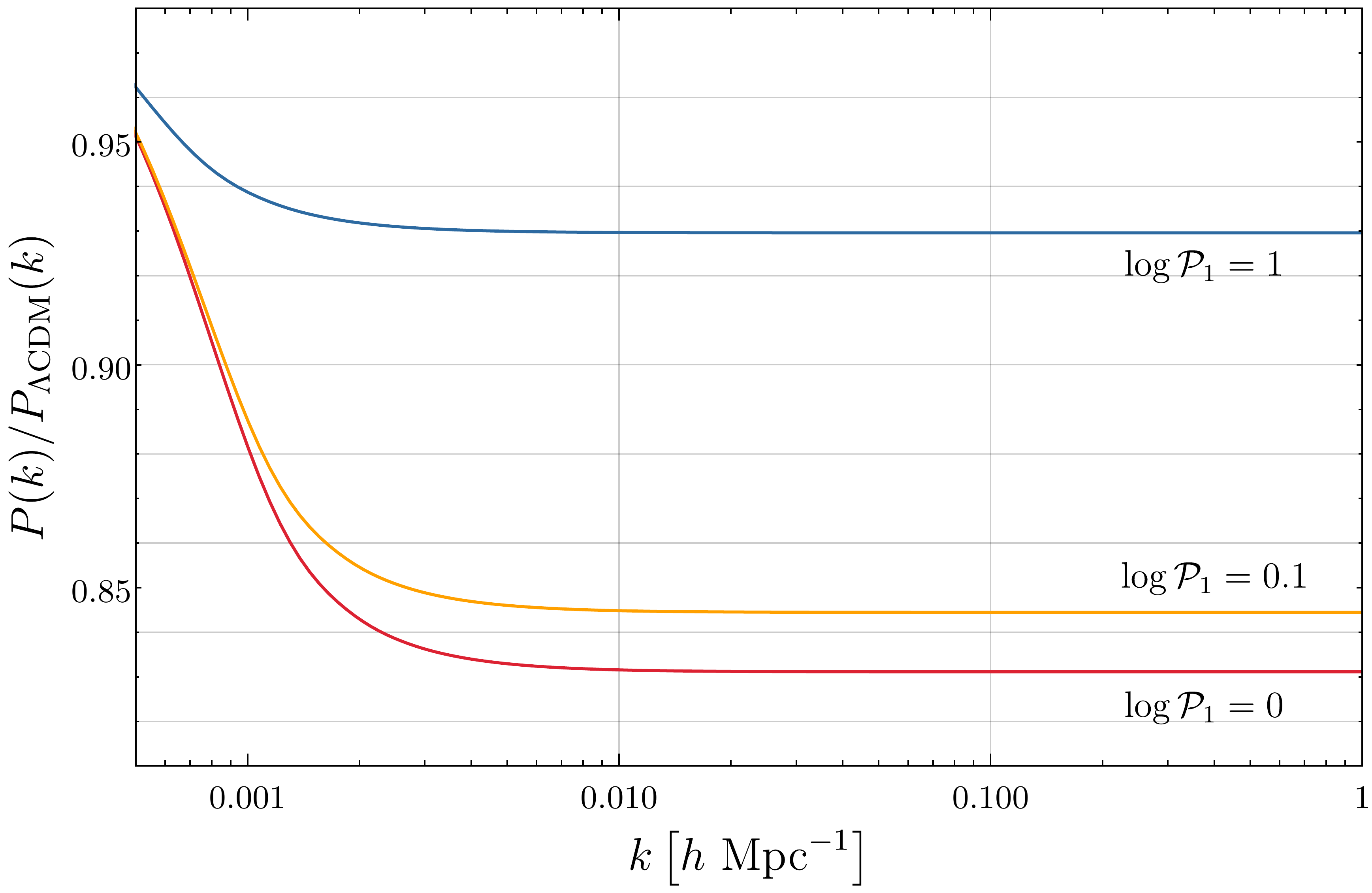} \includegraphics[width=0.495\textwidth]{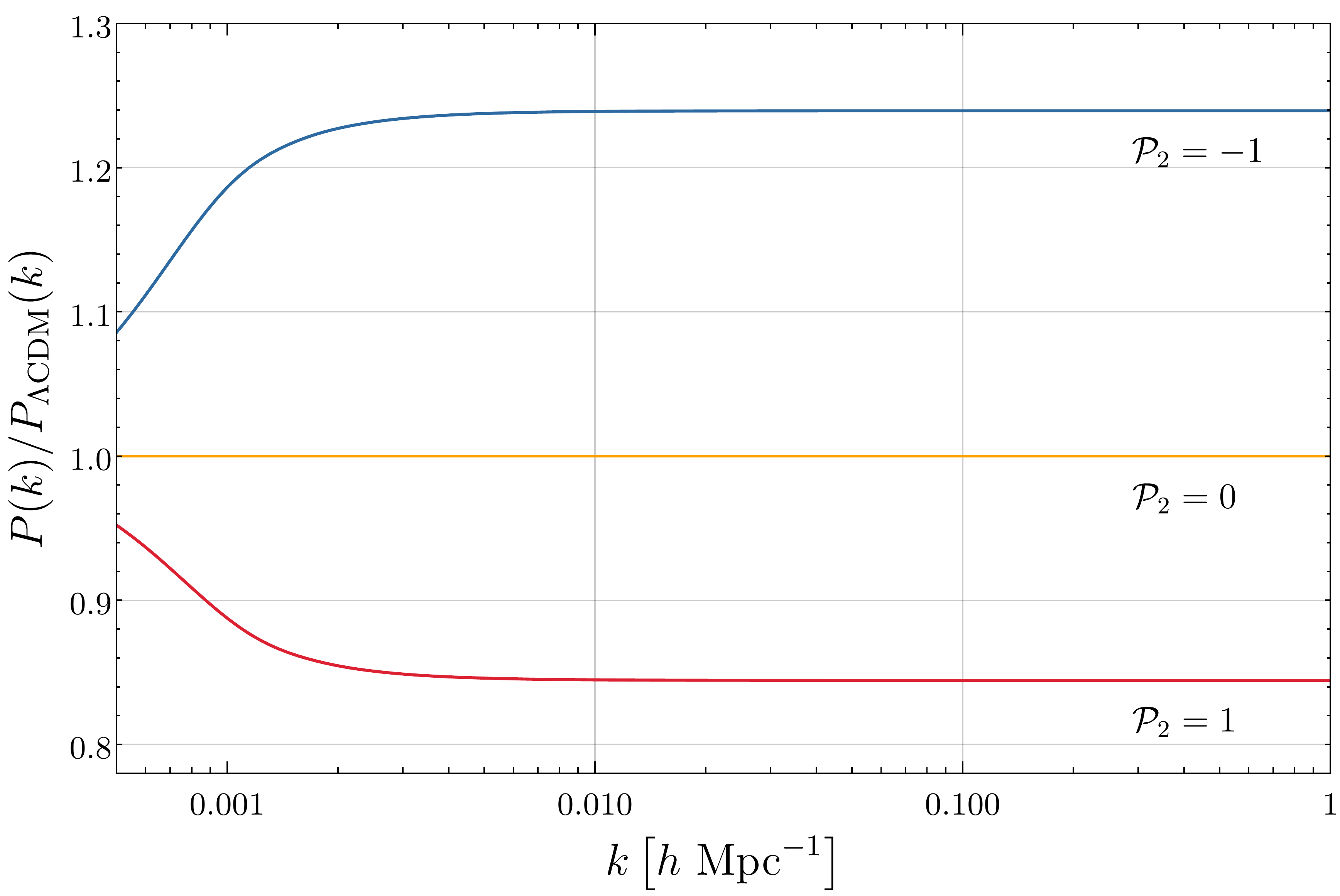} \\
 	\includegraphics[width=0.495\textwidth]{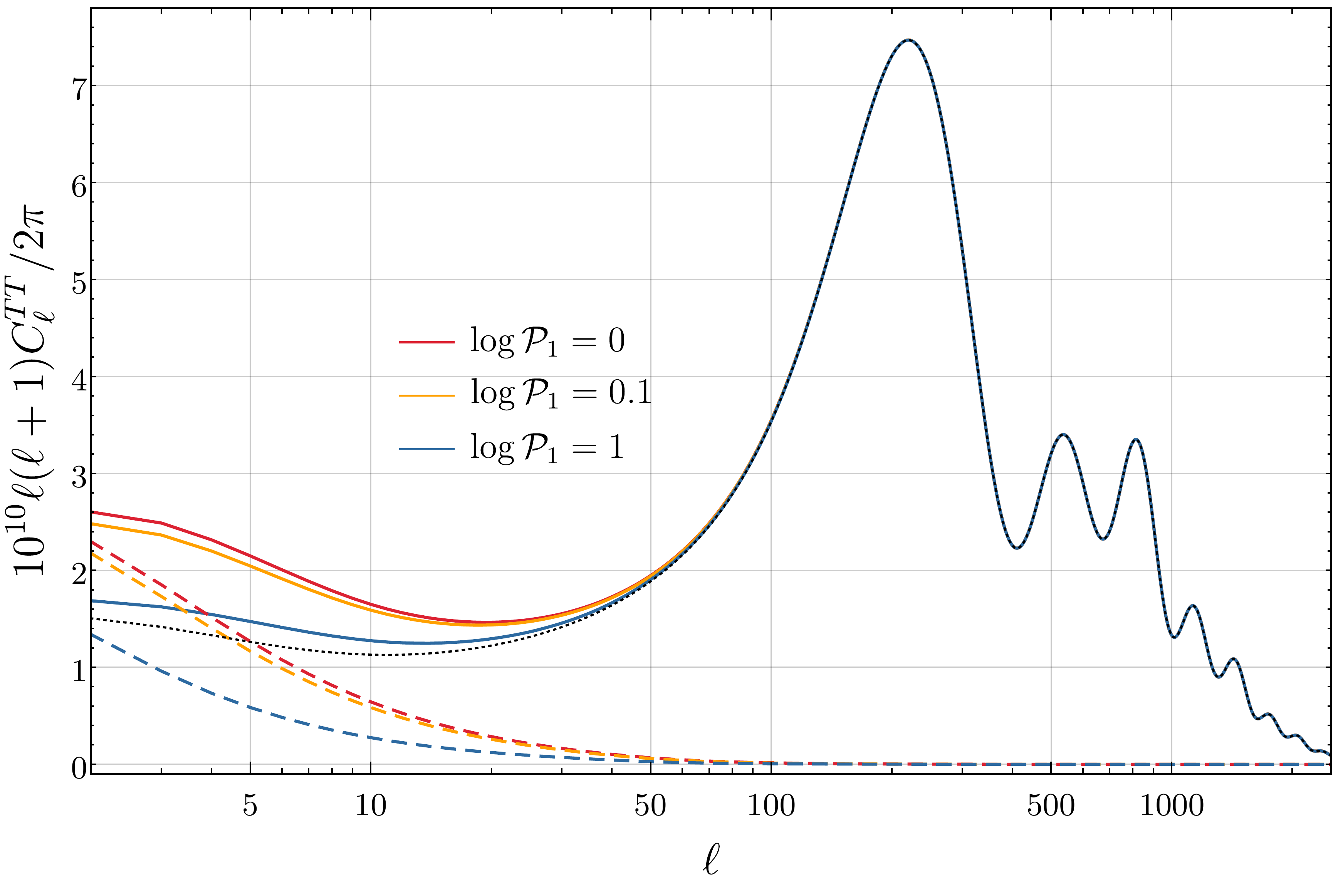} \includegraphics[width=0.495\textwidth]{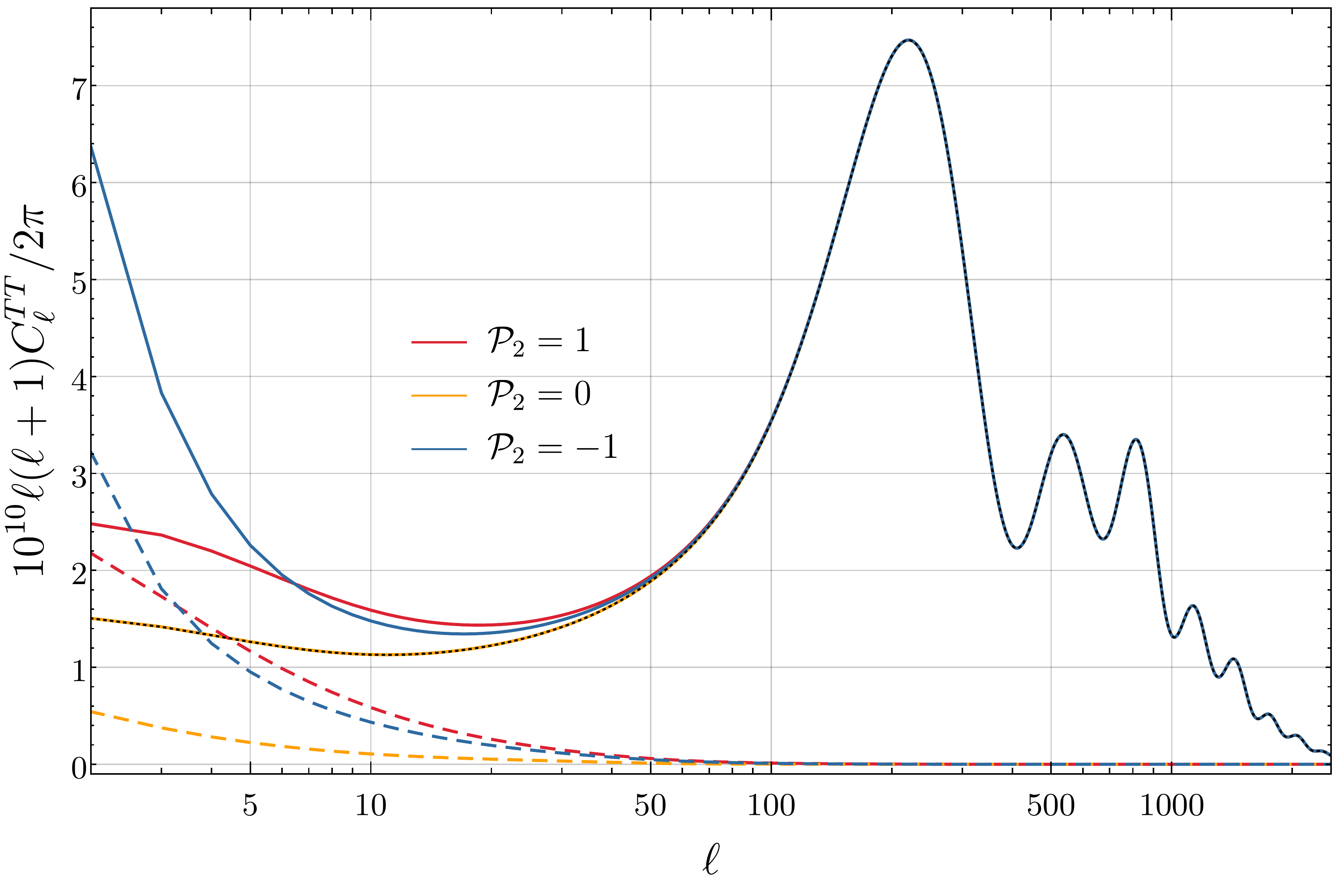}
 	\caption{\textit{Left panels}: The matter power spectrum relative to $\Lambda$CDM (top panel) and CMB temperature angular anisotropy power spectrum (bottom panel) for $\mathcal{P}_2=1$ and $w_{\rm de} = -1$ fixed. The late-time ISW component is shown by the dashed lines. The black dotted line denotes $\Lambda$CDM. \textit{Right panels}: As the left panels but with $\log \mathcal{P}_1=0.1$ fixed.}
 	\label{fig:Spectra}
 \end{figure*}
 
 \begin{figure*}
 	\centering
 	\includegraphics[width=0.495\textwidth]{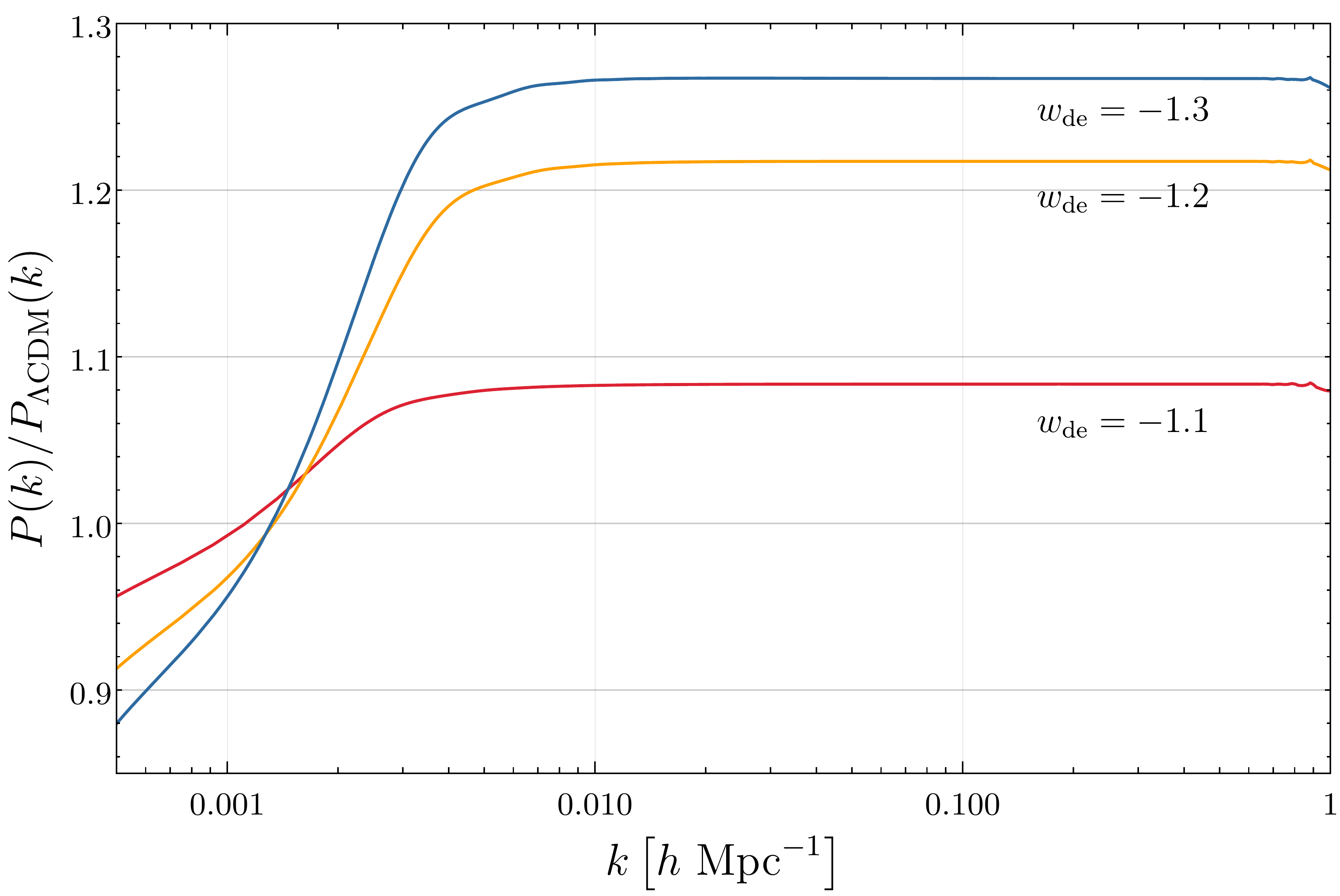} \includegraphics[width=0.495\textwidth]{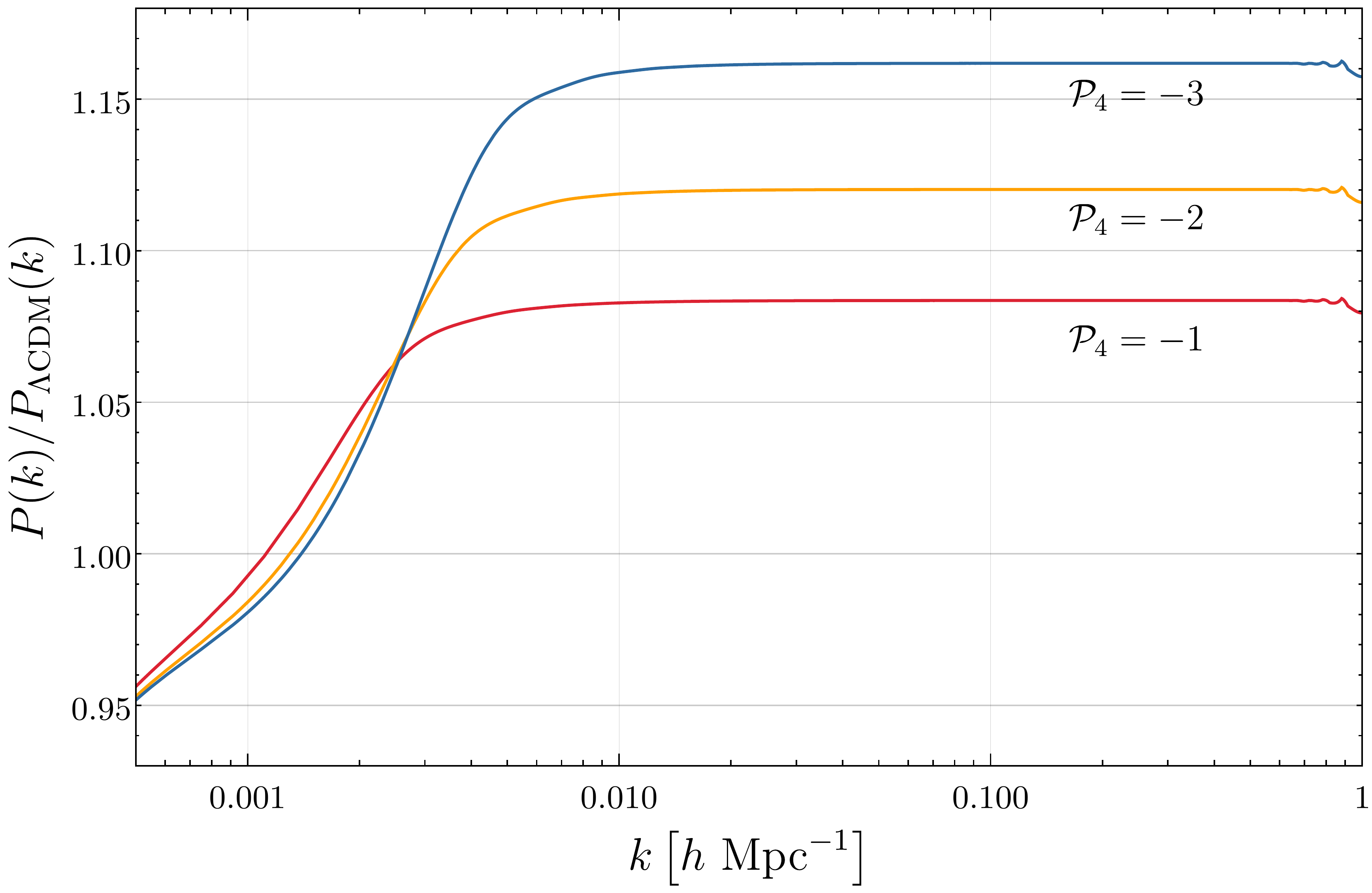} \\
 	\includegraphics[width=0.495\textwidth]{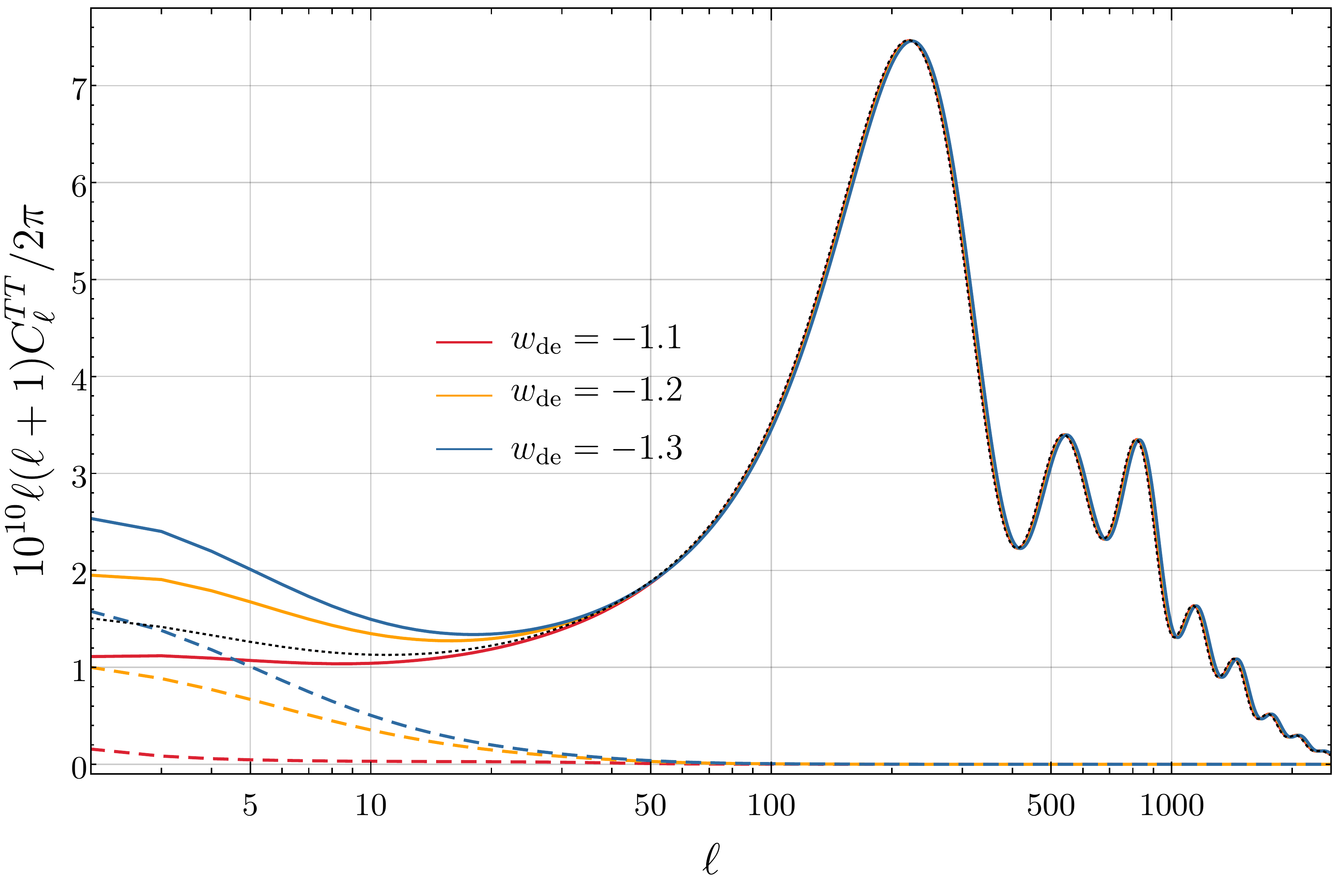} \includegraphics[width=0.495\textwidth]{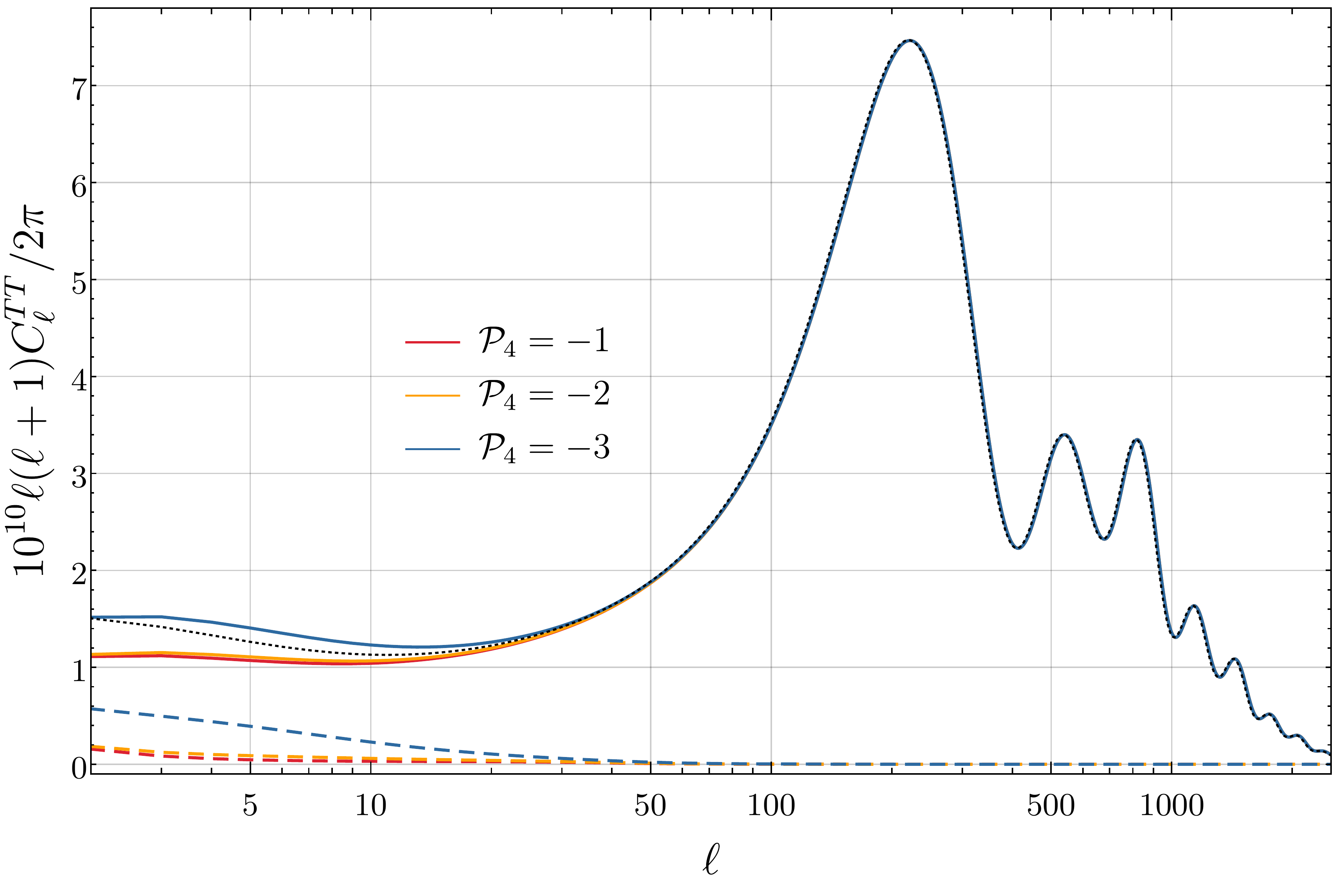}
 	\caption{\textit{Left panels}: The matter power spectrum relative to $\Lambda$CDM (top panel) and CMB temperature angular anisotropy power spectrum (bottom panel) in $w$CDM models, compatible with the gravitational waves constraint $c_{13}$ = 0, for different $w_{\rm de}$ and $\mathcal{P}_3=1$ and $\mathcal{P}_4=-1$ fixed. The late-time ISW component is shown by the dashed lines. The black dotted line denotes $\Lambda$CDM. \textit{Right panels}: As with the left panels, but for models with different $\mathcal{P}_4$ and $w_{\rm de} = -1.1$ and $\mathcal{P}_3=1$ fixed. All other cosmological parameters are fixed to $\Lambda$CDM values given previously.}
 	\label{fig:Spectra2}
 \end{figure*}
 
\section{Impact on cosmological observables} \label{sect:impact}
We now study the impact of designer $\F(\K)$ models on cosmological observables, and in particular on the CMB temperature angular anisotropy and matter power spectra. In doing so we will use the results derived in \autoref{sect:Evolve} and in particular $G_{\rm eff}$ which will directly affect the growth of matter and the Newtonian gravitational potentials.

\subsection{$w_{\rm de} = -1$}

 The impact of these models on the CMB temperature angular anisotropy power spectrum depends on the time variation of the Weyl potential, $2\Phi=\psi + \phi$. In particular, the late-time ISW effect is proportional to the integral $\int \dot{\Phi} \, dz$, for redshift $z$ along the line of sight. and so the presence of a non-zero $\Pi^{\rm S}_{\rm de}$ which will modify the behaviour of $\Phi$ will also affect the late-time ISW effect. Note that it is still possible for $\Phi$ to be different from $\Lambda$CDM even if $\Pi^{\rm S}_{\rm de}=0$, as long as $\Gamma_{\rm de} \not =0$. In the absence of $\Pi^{\rm S}_{\rm de}$ the gravitational slip $\eta = 1$ because $\psi = \phi$, but their behaviour will still be modified due to a non-zero $\Delta_{\rm de}$ in the Poisson equations for $\psi$ \eqref{eq:YEE} and $\phi$ \eqref{eq:ZEE}.
 
 From \autoref{fig:Phi}, we observe that for $G_{\rm eff,0}/G > 1$ $(< 1)$, $\Phi$ is enhanced (suppressed) with respect to $\Lambda$CDM. For $G_{\rm eff,0}/G > 1$ we see that $\Phi$ initially grows before decaying again due to dark energy. The growth can be explained from considering the matter power spectrum, since $G_{\rm eff,0}/G > 1$ would enhance clustering and hence give rise to a larger $\Phi$. In these cases we would expect the late-time ISW effect to be larger. For $G_{\rm eff,0}/G < 1$, $\Phi$ is suppressed relative to $\Lambda$CDM as expected and naively we may expect that the late-time ISW effect to be switched off for sufficiently low values of $G_{\rm eff,0}$. However, as seen in \autoref{fig:Spectra} we observe a larger late-time ISW effect relative to $\Lambda$CDM as $\dot{\Phi}$ is still larger in these cases as well, see \autoref{fig:Phi}. Therefore, there is a minimum in the late-time ISW effect for some value of $G_{\rm eff,0}$ before the late-time ISW effect grows again. Of course, it will not only depend on $G_{\rm eff,0}$ but also the overall behaviour of $G_{\rm eff}(a)$.
 
 Since $G_{\rm eff}$ also directly affects $\Delta_{\rm m}$ via \eqref{eq:2ODE DeltaM}, the matter power spectrum, $P(k)$, will be enhanced or suppressed according to the behaviour of $G_{\rm eff}$. Note that this will only be true for small scales as we have assumed we are in the regime ${\rm K} \gg 1$. Hence, designer $\F(\K)$ models will simultaneously modify the low-$\ell$ CMB temperature angular anisotropy power spectrum and $P(k)$ for large $k$. Indeed, this is what we find for $P(k)$ as shown in \autoref{fig:Spectra}. For large scales $P(k)$ is unaffected.
 
In principle there are 2 independent functions which are required to describe the behaviour of the Newtonian potentials, $\psi$ and $\phi$. What we call $G_{\rm eff}$ is sometimes referred to as $G_{\rm matter}$ or $\mu_{\phi}$, which parametrizes modifications to the Poisson equation for $\phi$, for example see \cite{Battye2018GW}. However there is an equivalent function which also modifies $\psi$, sometimes referred to as $G_{\rm light}$ or $\mu_{\psi}$. Therefore, to fully describe the behaviour of $\Phi = \psi + \phi$ we require knowledge of both $G_{\rm matter}$, what we call $G_{\rm eff}$, and $G_{\rm light}$. However, as mentioned previously, for designer $\F(\K)$ models with $w_{\rm de} = - 1$ we have that $\eta \rightarrow 1$ for ${\rm K} \gg 1$. This means that for subhorizon modes, where the expression for $G_{\rm eff}$ holds, the behaviour of $\psi$ and $\phi$ are identical. Therefore, only 1 function, $G_{\rm eff}$, is required to describe both the behaviour of $\psi$ and $\phi$. Note that this will also be true for our analysis of $w_{\rm de} \not= -1$ models as we set $c_{13} = 0$ which gives $\Pi_{\rm de}^{\rm S} = 0$ and so $\eta = 1$ at all scales.

  \begin{table*}[ht]
	\caption{The posterior mean (68\%C.L.) for $w_{\rm de}$, $\sigma_{8}$, $\log \mathcal{P}_1$, $\mathcal{P}_2$, $\mathcal{P}_3$, and $\mathcal{P}_4$ for 2 sets of models that mimic a $\Lambda$CDM and $w$CDM expansion history. The ellipses indicate parameters that are not used for that set of models. Note that for the $w$CDM models we study, $\mathcal{P}_1$ and $\mathcal{P}_2$ is set to zero and hence $\log \mathcal{P}_1 \rightarrow -\infty$.} \label{table:68 constraint}
	\begin{center}
		\renewcommand*{\arraystretch}{1.2}
		\begin{tabular}{|c||c|c||c|c|}
			\cline{2-5}
			\multicolumn{1}{c|}{} 
			& CMB ($\F(\K)$ $\Lambda$CDM) & CMB+Lensing ($\F(\K)$ $\Lambda$CDM) & CMB ($\F(\K)$ $w$CDM) & CMB+Lensing ($\F(\K)$ $w$CDM)\\
			\hline
			$w_{\rm de}$         & $-1$  & $-1$  & $-1.06^{+0.08}_{-0.03}$ & $-1.04^{+0.05}_{-0.02}$   \\
			\hline		
			$\sigma_{8}$         & $0.84\pm 0.02$ & $0.82\pm 0.01$ & $0.86^{+0.02}_{-0.03}$ & $0.83^{+0.01}_{-0.02}$ \\
			\hline					
			$\log{\mathcal{P}_1}$    & $1.7^{+2.3}_{-1.9}$ & $4.1^{+1.9}_{-1.3}$ & $-\infty$ & $-\infty$ \\
			\hline
			$\mathcal{P}_2$     & $-0.4\pm 0.5$ & $1.8^{+0.9}_{-3.2}$ & $0$ & $0$ \\
			\hline
			$\mathcal{P}_3$         & $\cdots$  & $\cdots$  & $1.3^{+17.0}_{-19.0}$ & $-0.7^{+19.9}_{-17.9}$ \\
			\hline
			$\mathcal{P}_4$         & $\cdots$  & $\cdots$  & $-1.7^{+1.2}_{-0.9}$ & $-1.0^{+1.0}_{-0.5}$  \\
			\hline
		\end{tabular}
	\end{center}
\end{table*}

\begin{table*}[ht]
	\caption{For comparison, the posterior mean (68\%C.L.) for $w_{\rm de}$, $\sigma_{8}$, for the standard $\Lambda$CDM and $w$CDM models.} \label{table:standard 68 constraint}
	\begin{center}
		\renewcommand*{\arraystretch}{1.2}
		\begin{tabular}{|c||c|c||c|c|}
			\cline{2-5}
			\multicolumn{1}{c|}{} 
			& CMB ($\Lambda$CDM) & CMB+Lensing ($\Lambda$CDM) & CMB ($w$CDM) & CMB+Lensing ($w$CDM)\\
			\hline
			$w_{\rm de}$         & $-1$  & $-1$  & $-1.54^{+0.18}_{-0.38}$ & $-1.36^{+0.31}_{-0.46}$   \\
			\hline		
			$\sigma_{8}$         & $0.83\pm 0.01$ & $0.82\pm 0.01$ & $0.98^{+0.11}_{-0.06}$ & $0.91^{+0.13}_{-0.03}$ \\
			\hline					
		\end{tabular}
	\end{center}
\end{table*}
 
 \begin{figure*}
 	\centering
 	\includegraphics[width=0.495\textwidth]{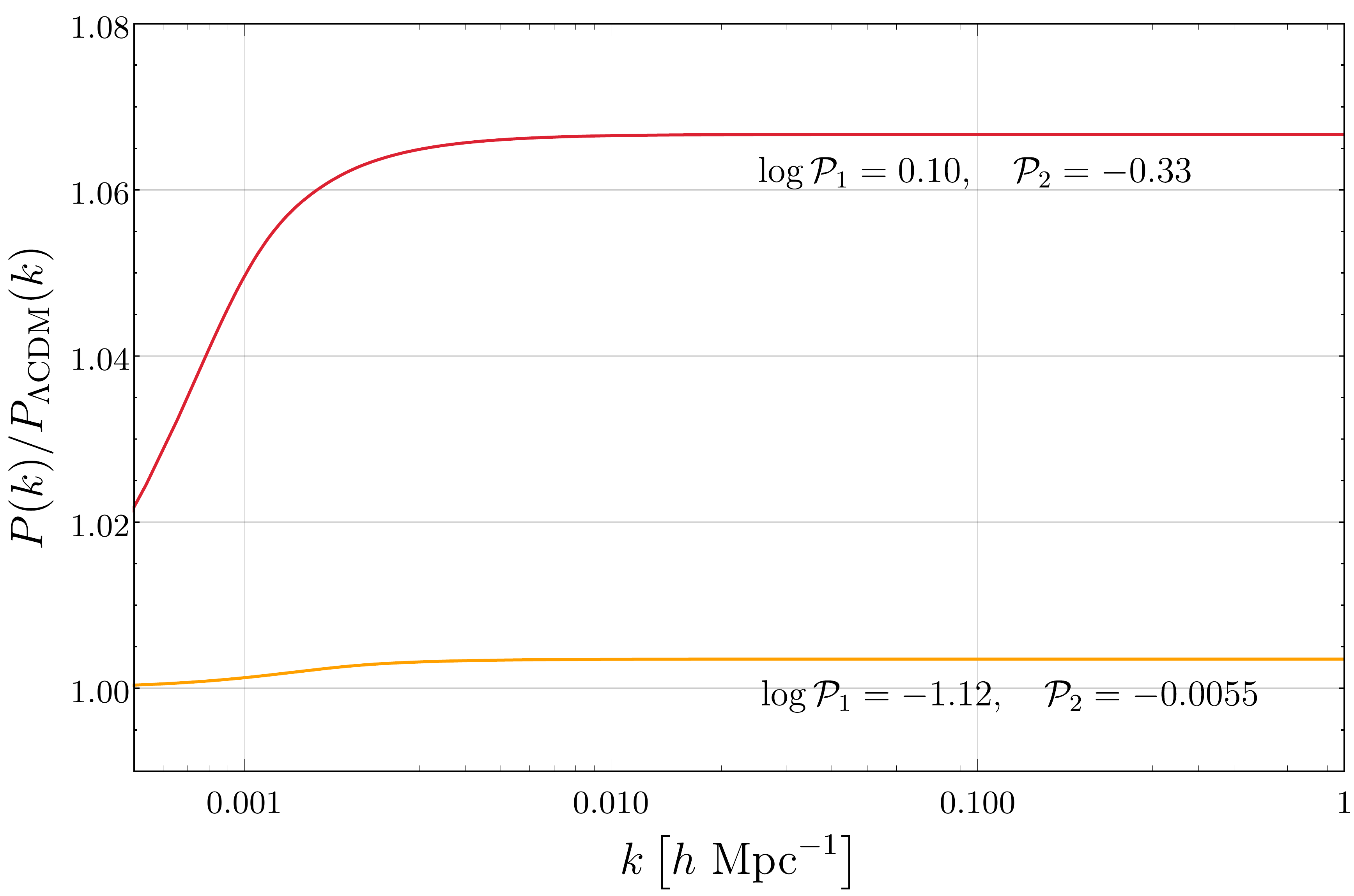} \includegraphics[width=0.495\textwidth]{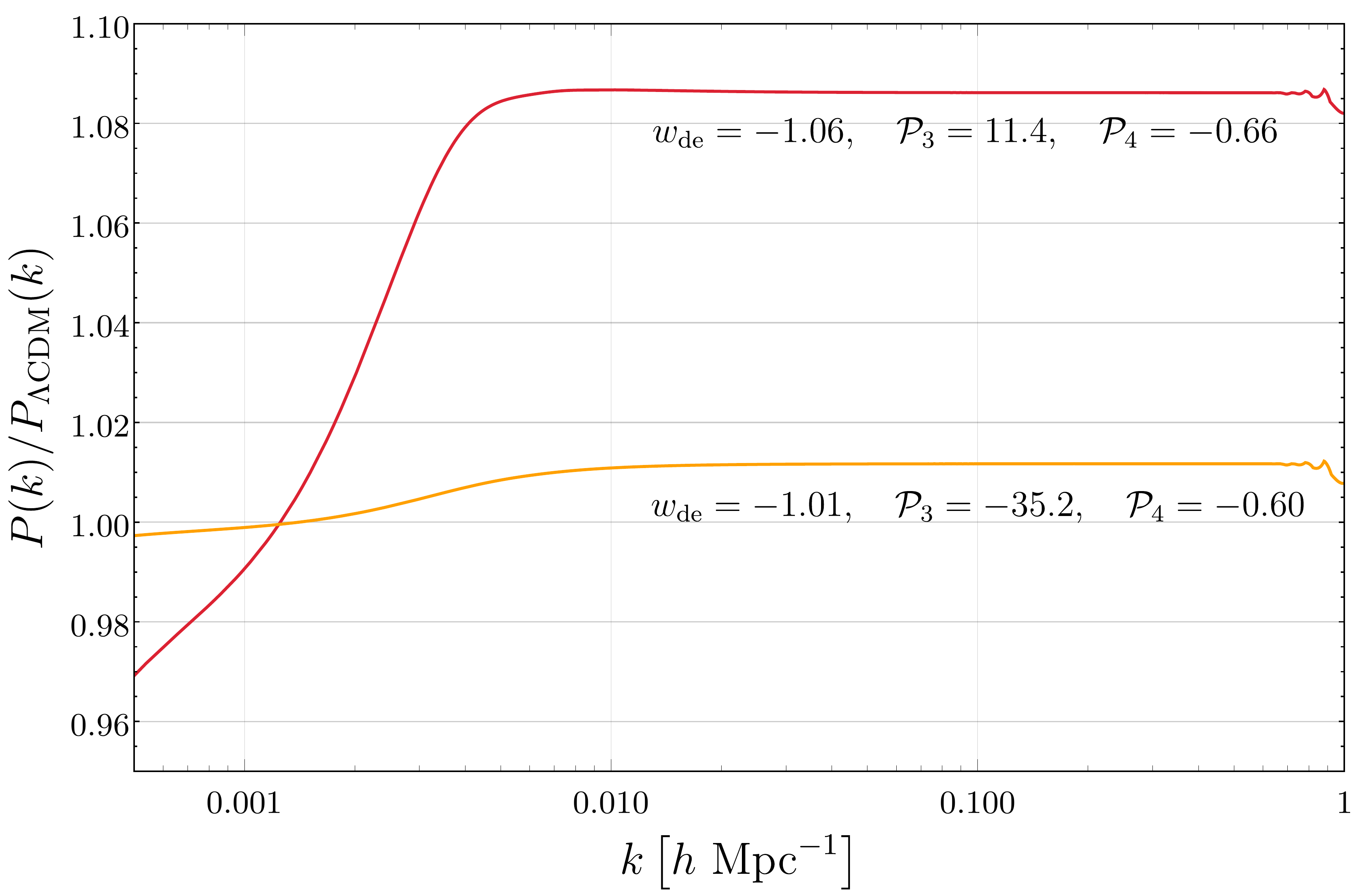} \\
 	\includegraphics[width=0.495\textwidth]{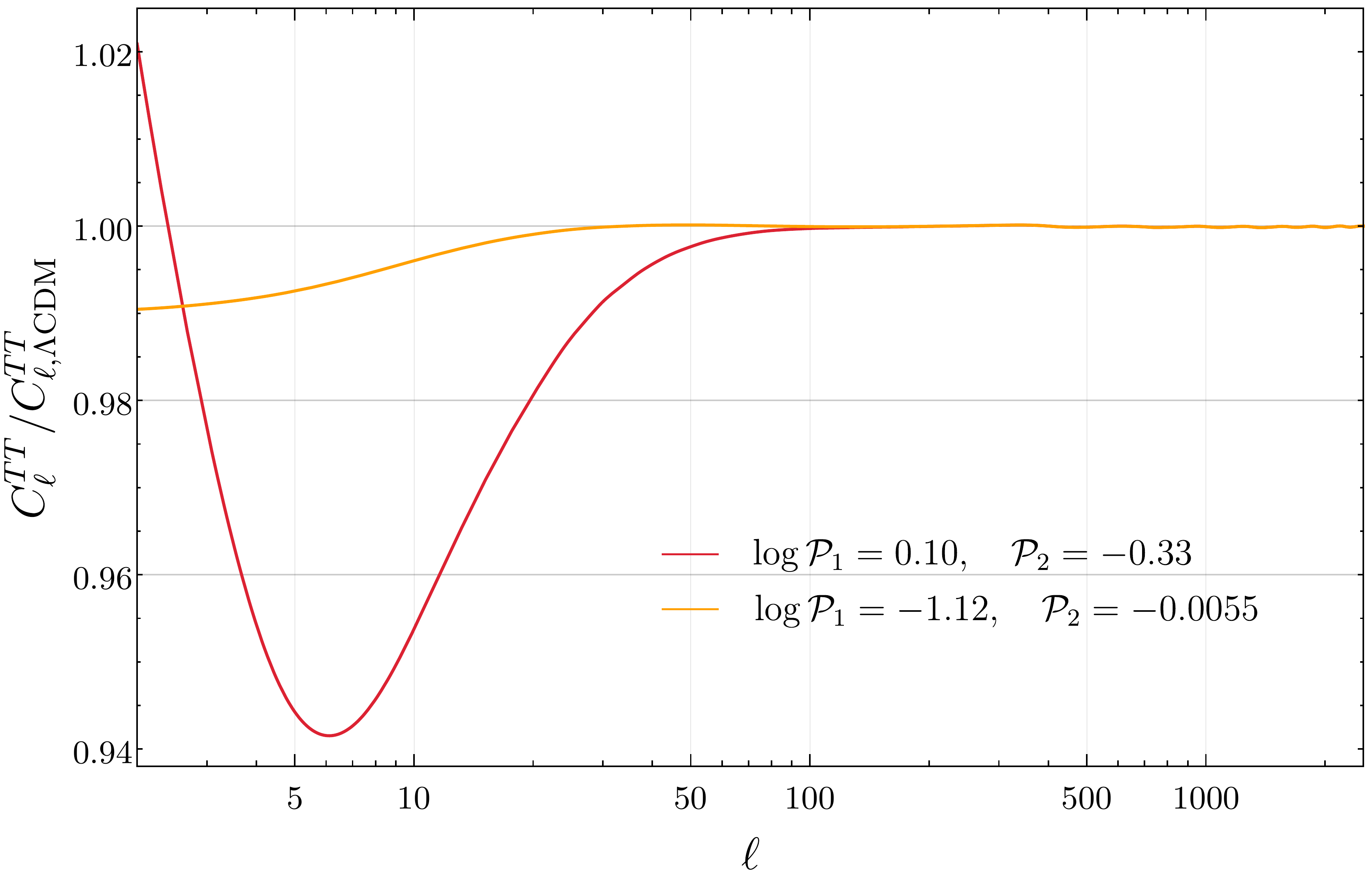} \includegraphics[width=0.495\textwidth]{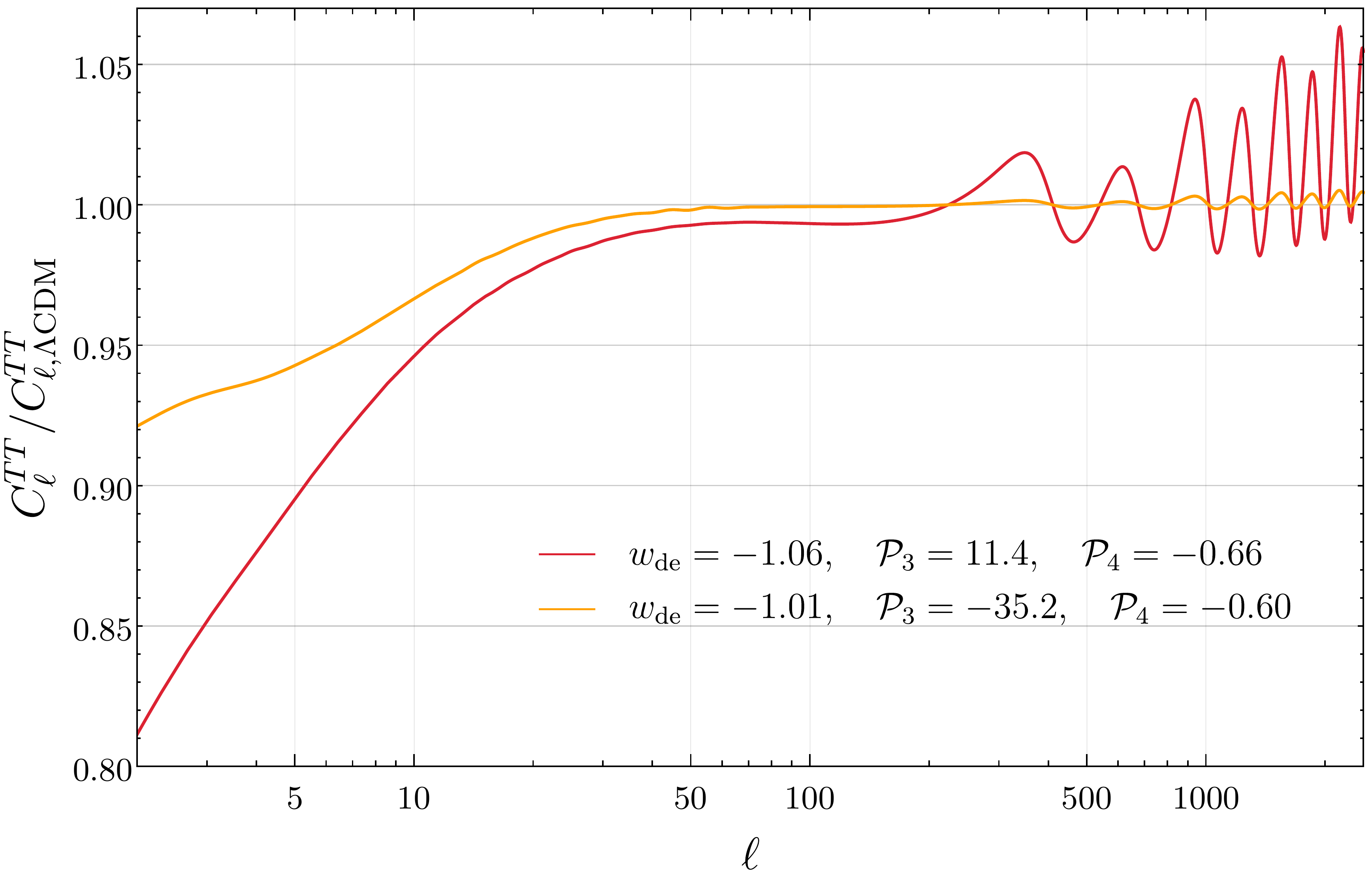}
 	\caption{\textit{Left panels}: The matter power spectrum (top panel) and CMB temperature angular anisotropy power spectrum (bottom) relative to $\Lambda$CDM for the best fitting values of $\log \mathcal{P}_1$ and $\mathcal{P}_2$, with $w_{\rm de} = -1$, for CMB (red) and CMB+Lensing (yellow). All other parameters have been kept fixed at the $\Lambda$CDM values. \textit{Right panels}: As the left panels, but for the best fitting values of $\mathcal{P}_3$, $\mathcal{P}_4$, and $w_{\rm de}\not= -1$, in $w$CDM models compatible with the gravitational waves constraint of $c_{13}=0.$ The oscillations observed at high-$\ell$ are due to a slightly different background cosmology which has shifted the peaks of the CMB spectrum.}
 	\label{fig:Best fit}
 \end{figure*}
 
 \subsection{$w_{\rm de} \not= -1$}

The behaviour of the spectra for $w_{\rm de} \not= -1$ is straightforward to understand. From \autoref{fig:Spectra2}, we see that $w_{\rm de}$ is anti-correlated with the amplitude of clustering, as in $w$CDM quintessence models \cite{Komatsu2009WMAP}. For more negative $w_{\rm de}$, dark energy domination begins later than more less negative values. Therefore, matter is more gravitationally bounded, enhancing the growth of structure. Similar to the $w_{\rm de} = -1$ models, this in turn affects the gravitational lensing potential, $\Phi$, enhancing the late-time ISW component of the CMB spectrum, as can be seen in \autoref{fig:Spectra2}.
 
As before with $\mathcal{P}_1$ and $\mathcal{P}_2$, $G_{\rm eff,0}$ \eqref{eq:Geff2} can be used to explain the impact of $\mathcal{P}_4$ on cosmological observables for $w_{\rm de} \not= -1$ models. We do not show the effect of varying $\mathcal{P}_3$ only as this is degenerate with $\mathcal{P}_4$ via \eqref{eq:P3P4}. The arguments that were made previously for $w_{\rm de} = -1$ also hold here. In fact, the behaviour for different $w_{\rm de}$ can also be explained via $G_{\rm eff}$. Even though the value of $G_{\rm eff, 0}$ in \eqref{eq:Geff2} is independent of $w_{\rm de}$, the behaviour of $G_{\rm eff}(a)$ is dependent on $w_{\rm de}$, with more negative values of $w_{\rm de}$ causing $G_{\rm eff}$ to be larger over its evolution, on average, compared to less negative $w_{\rm de}$.

It is interesting to note that in these models $\Pi^{\rm S}_{\rm de} = 0$ but we still observe differences in the late-time ISW effect. This suggests that even in models where $\Pi^{\rm S}_{\rm de} = 0$ and hence $\phi = \psi$, the late-time ISW effect can still be sensitive to a non-zero $\Gamma_{\rm de}$.

We will also briefly mention the effect of ignoring the $\left\lbrace \hat{\Theta}_i \right\rbrace$ terms in \eqref{eq:Gamma Einstein} for these models. From \eqref{eq:2ODE wDeltaDE}, it was possible to derive a second order equation of motion for $\Delta_{\rm de}$ coupled to $\Delta_{\rm m}$ assuming the $\left\lbrace \hat{\Theta}_i \right\rbrace$ terms were negligible in the ${\rm K} \gg 1$ regime. In these models, the coefficients of $\Gamma_{\rm de}$ written as \eqref{eq:Gamma Einstein} are given by
\begin{align}
c_{\Gamma\Delta_{\rm de}}&=\frac{6(1+w_{\rm de})}{f(\mathcal{P}_3,\mathcal{P}_4)\left[ 2\epsilon_H-3\Omega_{\rm de}(1+w_{\rm de})\right] } \\
&+\frac{\Omega_{\rm de}(1+w_{\rm de})(1+6w_{\rm de})-2w_{\rm de}\epsilon_H}{2\epsilon_H-3\Omega_{\rm de}(1+w_{\rm de})}, \\
c_{\Gamma\Theta_{\rm de}}&=\frac{4(\epsilon_H-1)\epsilon_H-2\epsilon_H'}{3\left[2\epsilon_H-3\Omega_{\rm de}(1+w_{\rm de}) \right] } \\
&-\frac{\Omega_{\rm de}(1+w_{\rm de})(1+3w_{\rm de})}{2\epsilon_H-3\Omega_{\rm de}(1+w_{\rm de})}, \\
c_{\Gamma\Delta_{\rm m}}&= - c_{\Gamma\Theta_{\rm m}} = \frac{1}{3}(1+w_{\rm de}),
\end{align} where we have defined $f(\mathcal{P}_3,\mathcal{P}_4) = a^{4+3w_{\rm de}}(\mathcal{P}_4-\mathcal{P}_3\beta_0)\frac{H}{H_0}+a\mathcal{P}_3\beta$ from \eqref{eq:wnot-1coeffs}. Since $w_{\rm de}$ will be close to $-1$, $c_{\Gamma\Theta_{\rm m}}$ will be close to zero and since the $\left\lbrace \Delta_i\right\rbrace $ terms will dominate anyway for ${\rm K} \gg 1$, a simplified model can be obtained by simply setting $c_{\Gamma\Theta_{\rm de}} = c_{\Gamma\Theta_{\rm m}} = 0$. In doing so we essentially ignore superhorizon modes, but it is interesting to see how much an effect this has on the spectra. We find that the error due to this simplification compared to the full model, when varying $w_{\rm de}$ and $\mathcal{P}_4$, is below $1\%$ down to $\ell \approx 10$ for $C_{\ell}^{\rm TT}$ and values of $k$ down to $k \approx 10^{-2} h \, {\rm Mpc}^{-1}$ for $P(k)$. As expected, for the largest scales, the spectra become increasingly inaccurate from ignoring the superhorizon modes in $\Gamma_{\rm de}$.

\section{Constraints from cosmological observables} \label{sect:Spectra}

In this section we present observational constraints for designer $\F(\K)$ models using CMB and CMB$+$Lensing data sets, from the \Planck 2015 public likelihoods for low-$\ell$, high-$\ell$ temperature with polarization, and lensing \cite{Planck2015XIII}. We do not consider data set combinations with baryon acoustic oscillations (BAOs), as these constrain only distance measurements and, as discussed in \cite{Battye2017GEA}, $\left\lbrace c_i, \F \right\rbrace $ would be insensitive to these, since they do not affect the background evolution. The caveat to this is that BAOs could give tighter constraints on other cosmological parameters which could then also affect the constraints on $\left\lbrace \mathcal{P}_i \right\rbrace$, however we do not consider this further. We reiterate that the main purpose of this analysis is not to get the best constraints for the parameters in this theory, but to understand how these models affect cosmological observables and if they can be used to solve some previously mentioned anomalies. 

For the MCMC sampling of the parameter space we use the \textsc{montepython} code \cite{Audren2013MP}. In this analysis, we vary the 6 base cosmological parameters and the required \Planck nuisance parameters with the same priors as the \Planck Collaboration. 

We consider 2 sets of models: one which mimics a $\Lambda$CDM background where $w_{\rm de} = -1$ and one which mimics $w$CDM. As discussed in \autoref{sect:EoS}, these two sets of models have different parameters to sample other than $w_{\rm de}$. The $68\%$C.L. constraints for these parameters are shown in \autoref{table:68 constraint}. For comparison we also show the $68\%$C.L. constraints we obtain for $w_{\rm de}$ and $\sigma_8$ in the standard cases, with the same data sets, in \autoref{table:standard 68 constraint}. Note that for $w$CDM, in this case, the parameters are poorly constrained due to the lack of BAO data which would constrain the background and hence $w_{\rm de}$. This is not the case for $\F(\K)$ models because the perturbations play a more important role than in the standard Quintessence case. Also note that we choose to explore over the parameter space for $\log \mathcal{P}_1$. As discussed previously, $\mathcal{P}_1$ is related to the sound speed, though there may be a caveat to this. Indeed we find that $\mathcal{P}_1$ must be positive otherwise the perturbations become unstable, and therefore $\log \mathcal{P}_1$ is a more suitable parameter.

For both sets of background cosmologies, we see that $\sigma_{8}$ decreases when lensing data is included. As seen previously, these models only affect the CMB temperature anisotropy on large angular scales, where the data are limited by cosmic variance. Hence, we expect the lensing power spectrum to more tightly constrain these models. Indeed, with the inclusion of lensing, models with larger $\sigma_{8}$ are disfavoured. For $w_{\rm de} = -1$, this pushes $\log \mathcal{P}_1$ higher, corresponding to a value of $G_{\rm eff,0}/G$ closer to 1 \eqref{eq:Geff}. As seen from \autoref{fig:contour1}, the higher values of $\log \mathcal{P}_1$ have weakened the constraints on $\mathcal{P}_2$ compared with CMB data only, though it is still consistent with zero. As mentioned previously, models with $\mathcal{P}_2 = 0$ are indistinguishable from $\Lambda$CDM at the perturbative level, which our constraints are consistent with. With $\mathcal{P}_1 \not= 0$, this also tells us that designer $\F(\K)$ models that mimic a $\Lambda$CDM background prefer $\F_\K = 0$, as opposed to $c_{13} = 0$, though these constraints from CMB data do not provide anything nearly as stringent as the gravitational waves constraint.

For models with $w_{\rm de} \not= -1$, compatible with $c_{13} = 0$, we find that a more negative value of $w_{\rm de}$ is preferred, though still consistent with $-1$ ($68\%$C.L.). The parameter $\mathcal{P}_3$ is poorly constrained by the data compared with $\mathcal{P}_4$. Including lensing data does little to improve this. It does, however, more tightly constrain $\mathcal{P}_4$ closer to zero, corresponding to a value of $G_{\rm eff,0}/G$ closer to 1 \eqref{eq:Geff2}. As mentioned previously this does not necessarily mean that $G_{\rm eff}(a) = 1$ for all $a$. Lensing data also more tightly constrains $w_{\rm de}$ pushing it closer to $-1$. The contour plot is shown in \autoref{fig:contour2}. It is worth noting that even with the gravitational waves constraint, which severely restricts many models, if $w_{\rm de} = -1$ is relaxed, then these designer $\F(\K)$ models are still able to be compatible with the data but are also distinct from $\Lambda$CDM due to the presence of a non-zero $\Gamma_{\rm de}$.

\begin{table*}[ht]
	\caption{The posterior mean (68\%C.L.) for $\sigma_{8}$, $\log \mathcal{P}_1$, $\mathcal{P}_2$, and $A_{\rm lens}$ in a designer $\F(\K)$ model with $w_{\rm de}=-1$ compared to the usual $\Lambda$CDM model.} \label{table:68 constraint Alens}
	\begin{center}
		\renewcommand*{\arraystretch}{1.2}
		\begin{tabular}{|c||c|c||c|c|}
			\cline{2-5}
			\multicolumn{1}{c|}{} 
			& CMB ($\F(\K)$ $\Lambda$CDM)& CMB+Lensing ($\F(\K)$ $\Lambda$CDM)& CMB ($\Lambda$CDM)& CMB+Lensing ($\Lambda$CDM) \\
			\hline		
			$\sigma_{8}$ & $0.80\pm0.02$ & $0.81\pm 0.01$ & $0.81^{+0.01}_{-0.02}$ & $0.81\pm0.01$ \\
			\hline					
			$A_{\rm lens}$     & $1.15^{+0.07}_{-0.08}$ & $1.12\pm0.05$ & $1.13\pm0.07$ & $1.12^{+0.05}_{-0.06}$\\
			\hline
			$\log{\mathcal{P}_1}$    & $5.4^{+2.1}_{-1.1}$ & $6.4^{+1.4}_{-0.7}$ & $\cdots$ & $\cdots$ \\
			\hline
			$\mathcal{P}_2$     & $28.9^{+9.0}_{-30.3}$ & $40.0^{+14.1}_{-41.3}$ & $\cdots$ & $\cdots$\\
			\hline 
		\end{tabular}
	\end{center}
\end{table*}

For $w_{\rm de} = -1$ models, the best fitting values for the parameters are $\log\mathcal{P}_1 = 0.10$, $\mathcal{P}_2 = -0.33$ (CMB), and $\log\mathcal{P}_1 = -1.12$, $\mathcal{P}_2 = -0.0055$ (CMB +Lensing). The spectra for these models are shown in \autoref{fig:Best fit}. We see that for CMB only, the best fitting parameters have the low-$\ell$ $C_l^{\rm TT}$ below that for the $\Lambda$CDM prediction, hence giving a better fit to the data. This is one of the current anomalies with the data mentioned previously \cite{Planck2016XX}. However, this comes at the cost of an enhanced matter power spectrum and hence a larger $\sigma_8$, at odds with galaxy clustering observations \cite{Torre2013VIMOS,Blake2012WZ,Blake2013galaxygrowth}, as with $f(R)$ models \cite{Battye2017FR}. With CMB+Lensing the conclusion is similar, but we find that the effects at the low-$\ell$ spectrum and high-$k$ matter power spectrum are much more subtle. We, therefore, find no strong reason to favour these models over $\Lambda$CDM for $w_{\rm de}=-1$ fixed.

For $w_{\rm de} \not= -1$ models, the best fitting values for the parameters are $w_{\rm de} =-1.06, \mathcal{P}_3 = 11.4$, $\mathcal{P}_4 = -0.66$ (CMB), and $w_{\rm de} =-1.01, \mathcal{P}_3 = -35.2$, $\mathcal{P}_4 = -0.60$ (CMB+Lensing). We see that, similar to $w_{\rm de} = -1$ models, the best fitting parameters have suppressed power at low-$\ell$ at the cost of a larger $\sigma_8$. The inclusion of lensing data causes the matter power spectrum to look more like $\Lambda$CDM, as before, however, there is slightly more suppression of power at low-$\ell$ compared to $w_{\rm de} = -1$ models.

 \begin{figure}
	\centering
	\includegraphics[width=0.495\textwidth]{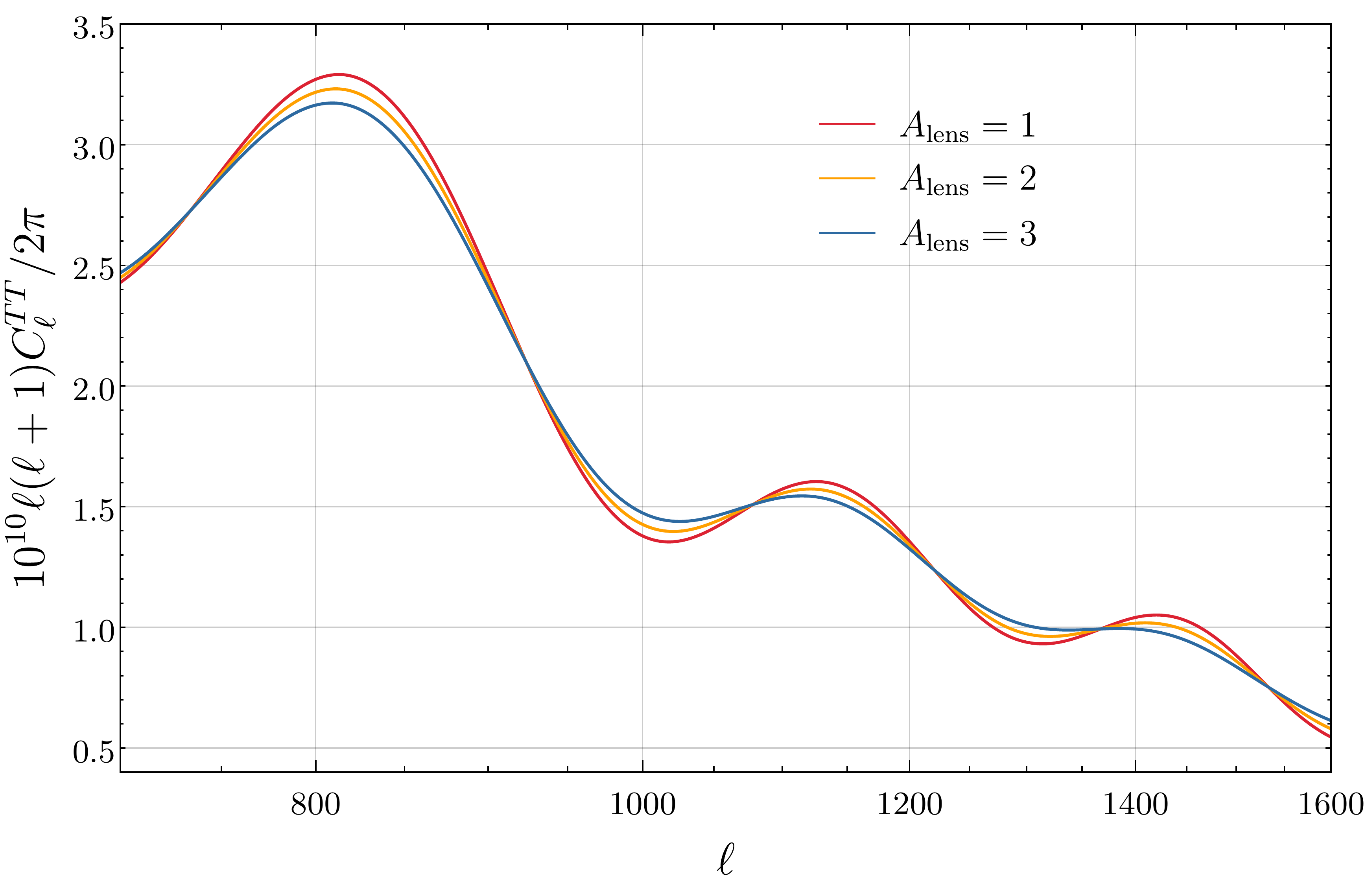} \\
	\includegraphics[width=0.495\textwidth]{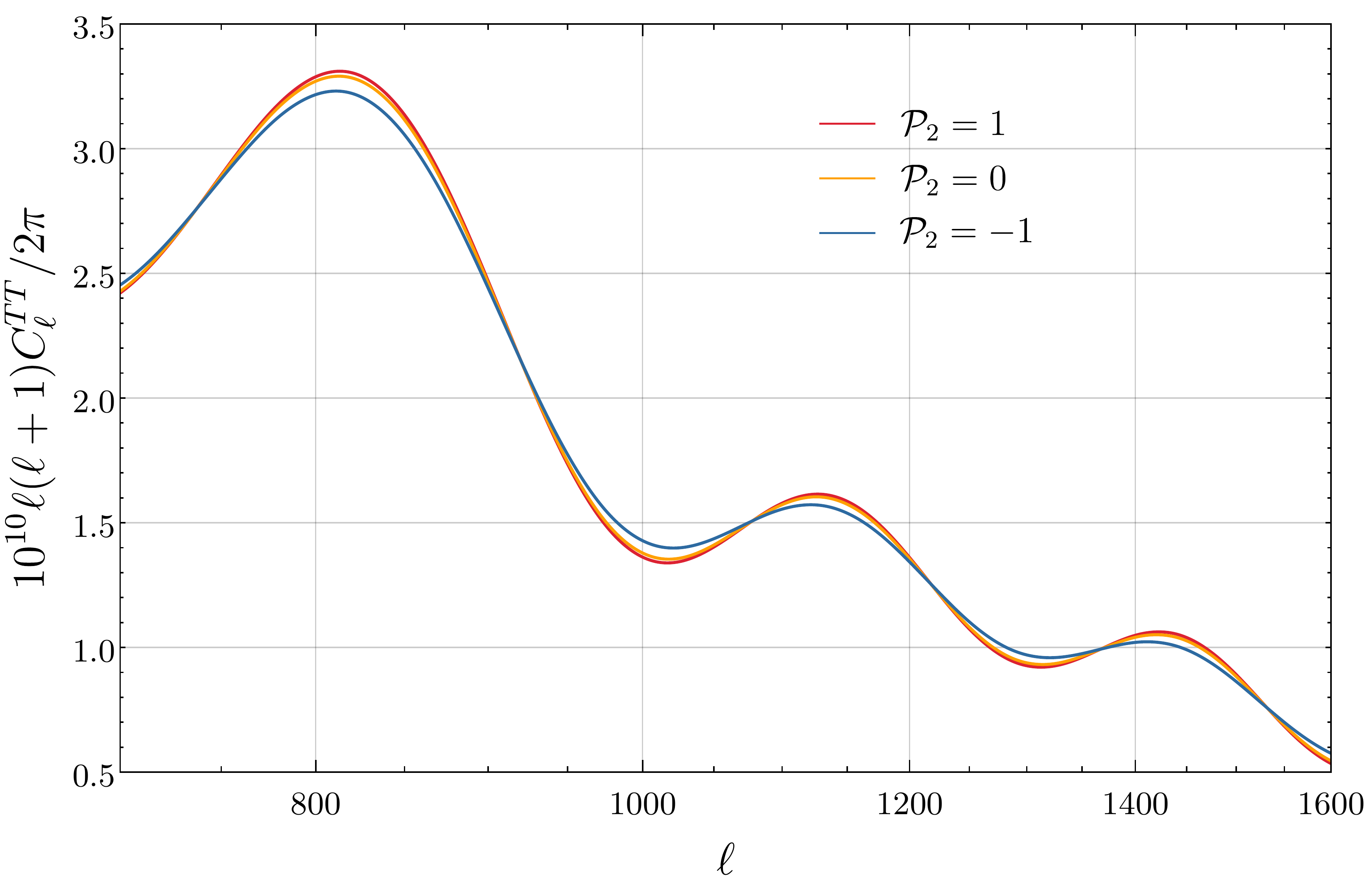} 
	\caption{\textit{Top panel}: The lensed CMB temperature angular anisotropy power spectrum, for high-$\ell$ peaks, in $\Lambda$CDM is shown. The amount of lensing is modified by $A_{\rm lens}$. A value of $A_{\rm lens} = 1$ is the expected amount of lensing. \textit{Bottom panel}: Similar to the top panel, but now in a designer $\F(\K)$ model with $w_{\rm de} = -1$. Here, $A_{\rm lens} = 1$ but the amount of lensing is affected by varying $\mathcal{P}_2$, with $\log \mathcal{P}_1 = -0.5$ fixed. }
	\label{fig:Lensing}
\end{figure}

\subsection{Gravitational lensing of the CMB power spectra}

As photons travel from the last scattering surface they are lensed from travelling through gravitational potentials. This gravitational lensing smooths the acoustic peaks of the CMB and polarization power spectra. The amount of lensing is something which can be calculated very accurately once the cosmological parameters are fixed \cite{Renzi2018Alens}. In \cite{Calabrese2008Alens} the parameter $A_{\rm lens}$ was introduced as a consistency check. This parameter modifies the amount the CMB is lensed via $C_\ell^{\rm TT, {\rm lensed}} = C_\ell^{\rm TT, {\rm unlensed}} + A_{\rm lens}\hat{C}_\ell^{\rm TT}$, where $\hat{C}_\ell^{\rm TT}$ is the lensed contribution to the unlensed spectrum. A theory which ignores gravitational lensing has $A_{\rm lens} = 0$, while $A_{\rm lens} = 1$ is the expected amount of lensing. We run the MCMC analysis on $\Lambda$CDM with $A_{\rm lens}$, shown in \autoref{table:68 constraint Alens}, and find that $A_{\rm lens} = 1$ is inconsistent ($68\%$C.L.), indicating more lensing is observed than expected in $\Lambda$CDM. This conclusion is compatible with previous analyses \cite{Calabrese2008Alens,Renzi2018Alens, Planck2016XLVI}.

In order to investigate this further, we also include the $A_{\rm lens}$ in our analysis of designer $\F(\K)$ models. Since this will modify the amount of gravitational lensing, a model which modifies the gravitational lensing potentials should also affect $A_{\rm lens}$. We have already seen that these designer $\F(\K)$ models can significantly modify $\Phi$ in \autoref{fig:Phi}. It is therefore interesting to see if the parameters in these models are degenerate with $A_{\rm lens}$ and hence could push $A_{\rm lens}$ to be more consistent with $1$. For illustrative purposes we will only consider models with $w_{\rm de} = -1$ together with $A_{\rm lens}$.

In \autoref{fig:Lensing} we show the high-$\ell$ peaks of the lensed CMB temperature angular anisotropy power spectrum. We see that the spectrum for higher values of $A_{\rm lens}$ have increasingly smoothed peaks. We compare this to a designer $\F(\K)$ model with $A_{\rm lens} = 1$ and see that a similar behaviour is observed by varying $\mathcal{P}_2$. Indeed, this is to be expected as this parameter, along with $\log \mathcal{P}_1$, directly affects the lensing potential $\Phi$. Prima facie, it seems that there should be a degeneracy between these new parameters and $A_{\rm lens}$, and that there may be a way to ameliorate the $A_{\rm lens}$ anomaly through $\mathcal{P}_1$ and $\mathcal{P}_2$. However, $A_{\rm lens}$ exclusively modifies the lensed high-$\ell$ CMB peaks, while $\mathcal{P}_1$ and $\mathcal{P}_2$ would affect both the high-$\ell$ peaks and low-$\ell$ late ISW effect, which would likely break the degeneracy. We again sample over the parameter space for designer $\F(\K)$ models with $w_{\rm de} = -1$, along with the same datasets as before. However, this time we also include the $A_{\rm lens}$ parameter. These constraints are shown in \autoref{table:68 constraint Alens}.

For CMB only, we find that $A_{\rm lens} = 1.15^{+0.07}_{-0.08}$ (CMB) and $A_{\rm lens} = 1.12\pm0.05$ (CMB+Lensing) at $68\%$C.L. Therefore, these models are not able to solve the $A_{\rm lens}$ anomaly. Indeed, there does not seem to be any degeneracy between $\left\lbrace \mathcal{P}_1,\mathcal{P}_2\right\rbrace$ and $A_{\rm lens}$, due to the fact that $\left\lbrace \mathcal{P}_1,\mathcal{P}_2\right\rbrace$ also modifies the low-$\ell$ CMB as shown in \autoref{fig:Spectra}. With $A_{\rm lens}$ we find that $\mathcal{P}_2$ becomes very poorly constrained. It is interesting to note that, unlike previously, the inclusion of lensing data pushes $\sigma_8$ slightly higher. This is due to the degeneracy with $A_{\rm lens}$ and $\sigma_8$ that can be seen in \autoref{fig:contour3}. We see that larger values of $A_{\rm lens}$ corresponds to lower $\sigma_8$. However, as the $C_\ell^{\varphi\varphi}$ spectrum does not exhibit the $A_{\rm lens}$ anomaly \cite{Planck2015XIII}, with the inclusion of lensing data, this pushes $A_{\rm lens}$ slightly closer to $1$, which in turn means that $\sigma_8$ is larger.

\section{Discussion and conclusions} \label{sect:Conclusion}

In this paper we have investigated cosmologically viable generalized Einstein-Aether theories, by which we mean models that are compatible with measurements of the expansion history of the Universe, data from the CMB photons, their polarization, and gravitational lensing potential, and also the gravitational waves constraint. In designer $\F(\K)$ models, the expansion history can be fixed to $w$CDM and the form of such an $\F(\K)$ for generic constant $w_{\rm de}$ was derived, given by \eqref{eq:AnalyticalFnotminus1}. These designer models are particularly useful for investigating the role of perturbations since they have the virtue that only the dynamics of the perturbations can be used to distinguish these models from $\Lambda$CDM or $w$CDM. 

To study the effect of such models on cosmological observables, we have used the EoS approach, implemented in a modified version of \textsc{class}, dubbed \textsc{class\textunderscore eos\textunderscore gea}. We have seen that a strength of this approach is that it has very readily identified the degeneracies that exist between the original parameters of the theory. This allows us to greatly reduce the number of parameters to explore in a MCMC analysis by constructing new parameters made from combinations of the previous ones, which are more suitable to explore over. In our case, the 5 original parameters $\left\lbrace c_i, \F_0 \right\rbrace $ could be reduced to 2, $\left\lbrace \mathcal{P}_1, \mathcal{P}_2\right\rbrace $  and $\left\lbrace \mathcal{P}_3, \mathcal{P}_4\right\rbrace $ for $w_{\rm de} = -1$ and $w_{\rm de} \not= -1$ with $c_{13} = 0$, respectively. Doing this is numerically more efficient and speeds up the computational analysis.

We found that for designer $\F(\K)$ models with $w_{\rm de}=-1$, the data seems to prefer models with a small derivative, i.e. $\F_\K \approx 0$, corresponding to $\mathcal{P}_2 \approx 0$. Such models are consistent with the gravitational waves constraint, but are also indistinguishable from $\Lambda$CDM. The other way to satisfy the constraint is to have $c_{13}=0$, however this was also shown to cause the models to be indistinguishable from $\Lambda$CDM. While there exists a choice of parameters which would suppress power at low-$\ell$ for the CMB temperature angular anisotropy power spectrum, see \autoref{fig:Best fit}, this came at the cost of an increased $\sigma_8$. Moreover, these effects were diminished with the inclusion of lensing data. This is due to the lensing data disfavouring models with large values of $\sigma_8$. Therefore, these models do not provide a significant alternative to $\Lambda$CDM.

Since $c_{13} = 0$ causes the previous set of models to be indistinguishable from $\Lambda$CDM, to explore cosmologically viable, but also interesting models, the case of $w_{\rm de}\not= -1$ was investigated. We found that in such models, $w_{\rm de}$ was constrained to be $w_{\rm de } = -1.07^{+0.08}_{-0.03}$ (CMB) and $w_{\rm de } =-1.04^{+0.05}_{-0.01}$ (CMB+Lensing) at $68\%$C.L. Since $w_{\rm de }$ is anti-correlated with $\sigma_8$ it is not surprising that the value of $w_{\rm de}$ is pushed closer to $-1$ with the inclusion of lensing data. We find $w_{\rm de} = -1$ to be consistent, i.e. these models need to be close to $\Lambda$CDM in order to be compatible with the data. However, they do not need to be exactly $\Lambda$CDM and there is some leeway for these models to fit the data but to also have noticeable differences. In particular, the gravitational waves constraint does not severely restrict these models since those constraints pertain only to those with significant $\Pi_{\rm de}^{\rm S}$, which these models avoid, since the constraints on $\Gamma_{\rm de}$ are much weaker. Similar to before, there exists a choice of parameters to suppress power for the low-$\ell$ CMB temperature angular anisotropy power spectrum, but at the cost of a larger $\sigma_8$. Again, while these models are in principle cosmologically viable, we do not see any reason to favour these models over $\Lambda$CDM. However, it is interesting to note that some of the anomalies with $\Lambda$CDM can be rectified in these alternative dark energy models, although not simultaneously. For example, it is possible to suppress power at high-$k$ for $P(k)$, as shown in \autoref{fig:Spectra}, but at the cost of enhanced power for the low-$\ell$ CMB spectrum, and vice-versa.

When investigating the $A_{\rm lens}$ anomaly within $\F(\K)$, we found comparable constraints on $A_{\rm lens}$ with previous analyses, i.e. $A_{\rm lens} = 1$ is not consistent in these models. Since the data suggests that these models need to be close to $\Lambda$CDM in order to be cosmologically viable, this is not surprising. It is currently unclear whether these previously mentioned anomalies are due to unaccounted systematics in the data, or whether there is new physics to be understood. It may be possible to construct models which are able to simultaneously alleviate these anomalies, i.e. low-$\ell$ CMB, high-$k$ matter power spectrum, and the $A_{\rm lens}$ anomaly. As we have seen, these can be linked to the Weyl potential, $\Phi$, which is affected by $G_{\rm eff}$ only in our case, though 2 functions are required in general when $\Pi^{\rm S} \not =0$. Therefore, it may be possible to construct models with a suitable $G_{\rm eff}$ by choosing $\left\lbrace c_{\Pi}, c_{\Gamma} \right\rbrace $, which then could be used to investigate these anomalies further and see what properties models would need in order for these anomalies to be solved in some way. We leave this for future work.

\section*{Acknowledgements}

DT is supported by an STFC studentship. RAB and FP acknowledge support from STFC grant ST/P000649/1. BB acknowledges financial support from the ERC Consolidator Grant 725456. Part of the analysis presented here is based on observations obtained with \Planck (http://www.esa.int/Planck), an ESA science mission with instruments and contributions directly funded by ESA Member States, NASA, and Canada.

\appendix

\section{Coefficients in the Equations of State approach} \label{app:A}

The coefficients for the equations of state in generalized Einstein-Aether theories are presented here. From \eqref{eq:Pi2} and \eqref{eq:Gamma2}, we have that
\begin{align} \label{c_PiD}
c_{\Pi \Delta} &= \frac{c_{13}}{c_{14}}, \\
c_{\Pi \Theta} &= \frac{c_{13}}{3c_{123}+2\alpha \gamma_2}\left[1- 2\left( \epsilon_H \gamma_2+ \frac{c_{13}}{c_{14}}\right) \right] ,   \\
c_{\Pi X}&= \frac{2c_{13}\gamma_1(1+2\gamma_2)}{\left( 2\gamma_1-1 \right) (3c_{123}+2\alpha \gamma_2)}\left[ 2\left( \frac{c_{13}}{c_{14}} +\epsilon_H\gamma_2\right) -1 \right],  \\ \label{c_PiY}
c_{\Pi Y} &= \frac{2c_{13}\gamma_1}{3\alpha \left(1-2\gamma_1 \right) }, \\ \label{c_GammaD}
c_{\Gamma\Delta} &= \frac{\alpha (1+2\gamma_2)}{3c_{14}} - \frac{dP_{\rm de}}{d\rho_{\rm de}}, \\
c_{\Gamma\Theta} &= \frac{\alpha}{3 (3c_{123}+2\alpha \gamma_2)}\left[\left( 1-\frac{2c_{13}}{c_{14}}\right) \right. \nonumber \\ &\left. (1+2\gamma_2)  - 6\epsilon_H \gamma_2\left( 1+\frac{2}{3}\gamma_3\right)   \right] + \frac{dP_{\rm de}}{d\rho_{\rm de}}, \\
c_{\Gamma W} &= \frac{2\gamma_1(1+2\gamma_2)}{3\left( 2\gamma_1-1 \right) }, \\
c_{\Gamma X}  &= \frac{4\alpha\gamma_1}{3\left( 2\gamma_1-1 \right)(3c_{123}+2\alpha \gamma_2)}\left[\left(1+ \frac{c_{13}}{c_{14}}\right) (1+2\gamma_2)^2 \right.\nonumber \\ &\left.+ \frac{3c_{13}}{\alpha}\left(1+2\gamma_2\left[1-\epsilon_H\left(1+\frac{2}{3}\gamma_3 \right)  \right] \right) \right], \\ \label{c_GammaY}
c_{\Gamma Y} &= \frac{2\gamma_1(1+2\gamma_2)}{9\left( 1-2\gamma_1\right) },
\end{align} where $\gamma_1 = \K\F_\K/\F$ , $\gamma_2 = \K\F_{\K\K}/\F_{\K}$, and $\gamma_3 = \K\F_{\K\K\K}/\F_{\K\K}$. Using the Einstein equations \eqref{eq:WEE} - \eqref{eq:ZEE} to eliminate the metric variables $\left\lbrace W,X,Y\right\rbrace$, we obtain
\begin{align} \label{Pi DM}
(1-&3c_{\Pi Y}\Omega_{\mathrm{de}})w_\mathrm{de}\Pi^{\rm S}_\mathrm{de} \nonumber \\= &\left( c_{\Pi\Delta}-\frac{3}{2}c_{\Pi Y}\Omega_\mathrm{de}\right) \Delta_\mathrm{de}+\left(c_{\Pi\Theta}+\frac{1}{2}c_{\Pi X}\Omega_\mathrm{de} \right) \hat{\Theta}_\mathrm{de} \nonumber\\
&-\frac{3}{2}c_{\Pi Y}\Omega_\mathrm{m}\Delta_\mathrm{m}+\frac{1}{2}c_{\Pi X}\Omega_\mathrm{m}\hat{\Theta}_\mathrm{m}+3c_{\Pi Y}\Omega_\mathrm{m}w_\mathrm{m}\Pi^{\rm S}_\mathrm{m},
\end{align}
\begin{align} \label{Gamma DM}
\frac{1}{2}( &2-3c_{\Gamma W}\Omega_\mathrm{de})w_\mathrm{de}\Gamma_\mathrm{de} \nonumber \\ = &\left( c_{\Gamma \Delta}+\frac{3}{2}c_{\Gamma W}\Omega_\mathrm{de}\left. \frac{dP}{d\rho}\right|_\mathrm{de}-\frac{3}{2}c_{\Gamma Y}\Omega_\mathrm{de} \right) \Delta_\mathrm{de} \nonumber \\ &+\frac{3}{2}\Omega_\mathrm{m}\left(c_{\Gamma W} \left. \frac{dP}{d\rho}\right|_\mathrm{m} -c_{\Gamma Y}\right)\Delta_\mathrm{m} \nonumber \\ &+\left[ c_{\Gamma \Theta}-\frac{3}{2}c_{\Gamma W}\Omega_\mathrm{de}\left( 1 + \left. \frac{dP}{d\rho}\right|_\mathrm{de} \right)+\frac{1}{2}c_{\Gamma X}\Omega_\mathrm{de}  \right]\hat{\Theta}_\mathrm{de} \nonumber \\
&+\frac{1}{2}\left[c_{\Gamma X}-3c_{\Gamma W} \left( 1 + \left. \frac{dP}{d\rho}\right|_\mathrm{m} \right)  \right]  \Omega_\mathrm{m}\hat{\Theta}_\mathrm{m} \nonumber \\ &+\frac{3}{2}c_{\Gamma W}\Omega_\mathrm{m}w_\mathrm{m}\Gamma_\mathrm{m}.
\end{align} 

\clearpage
\onecolumngrid
\section{Contour plots for designer $\F(\K)$ parameters} \label{app:B}
\begin{figure*}[h]
	\centering
	\includegraphics[width=\textwidth]{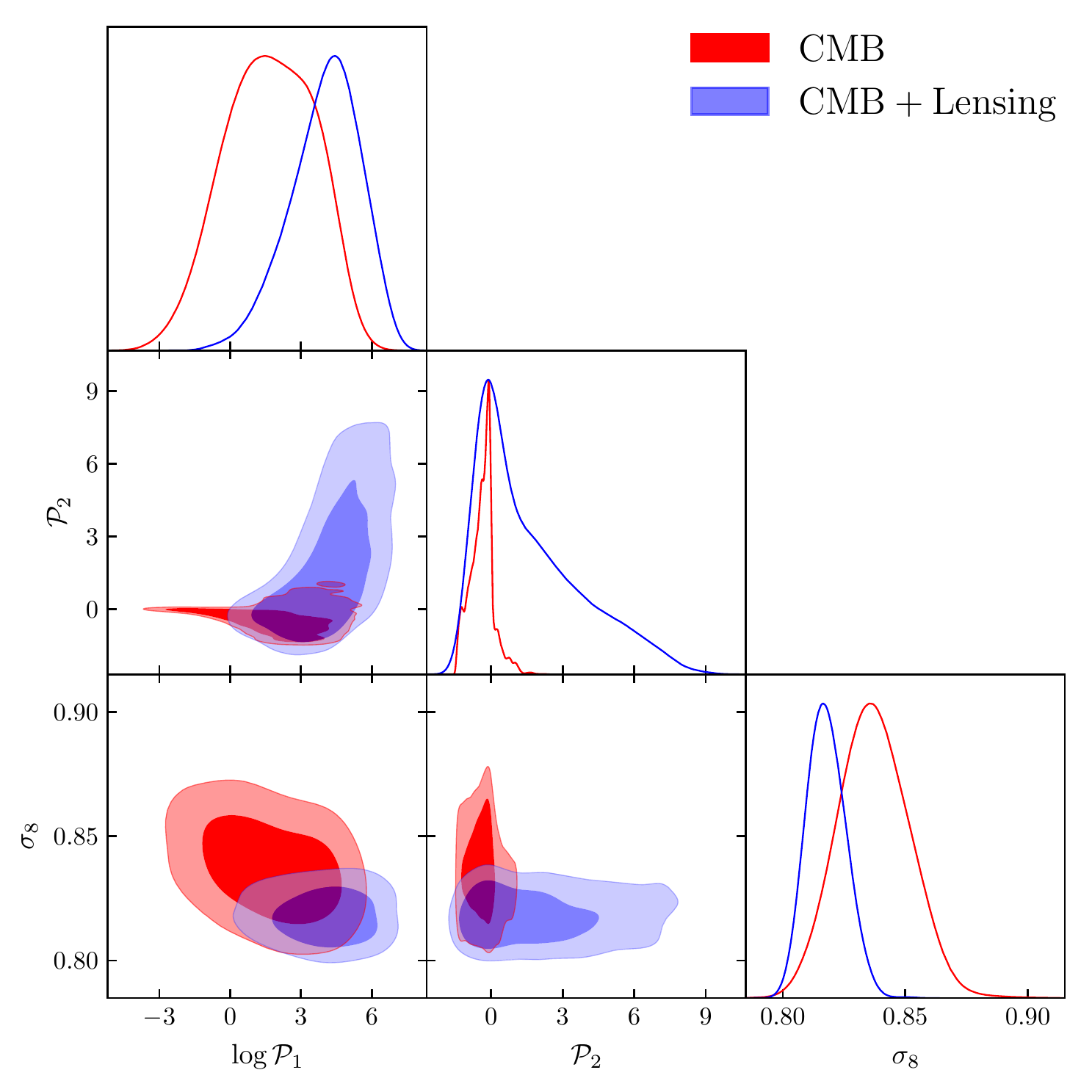} 
	\caption{The $68\%$ and $95\%$ constraint contours for $\log \mathcal{P}_1$, $\mathcal{P}_2$, and $\sigma_8$ are shown for $w_{\rm de} = -1$ models. Note that the correlation between $\log \mathcal{P}_1$ and $\sigma_8$ is removed once lensing data is included.} \label{fig:contour1}
\end{figure*}

\begin{figure*}[h]
	\centering
	\includegraphics[width=\textwidth]{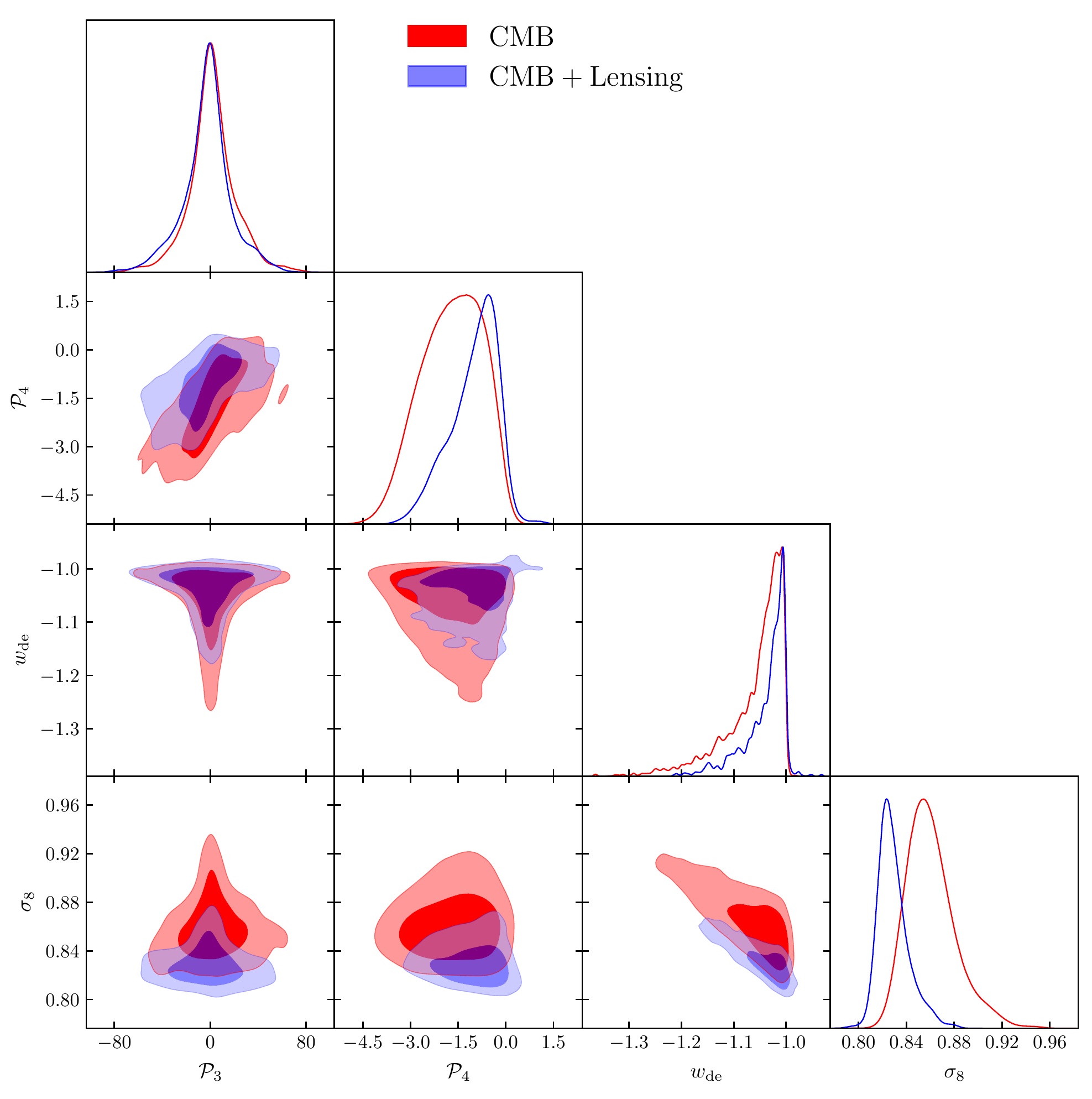} 
	\caption{The $68\%$ and $95\%$ constraint contours for $\mathcal{P}_3$, $\mathcal{P}_4$, $\sigma_8$, and $w_{\rm de}$. Note the anti-correlation between $\sigma_8$ and $w_{\rm de}$ as in $w$CDM quintessence models. As expected, there is a degeneracy between $\mathcal{P}_3$ and $\mathcal{P}_4$.} \label{fig:contour2}
\end{figure*}

\begin{figure*}[h]
	\centering
	\includegraphics[width=\textwidth]{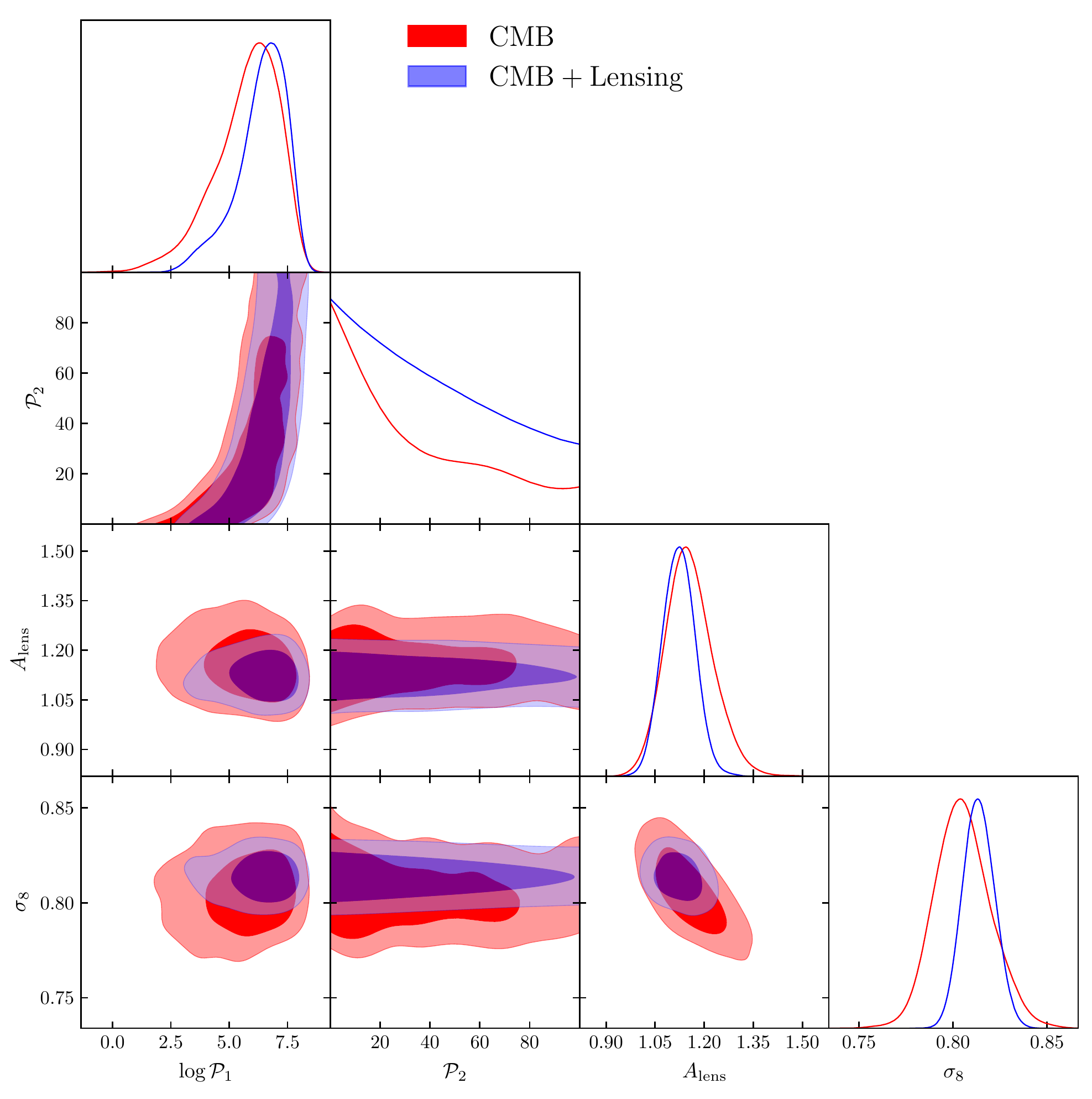} 
	\caption{The $68\%$ and $95\%$ constraint contours for $\log \mathcal{P}_1$, $\mathcal{P}_2$, $A_{\rm lens}$ and $\sigma_8$ are shown for $w_{\rm de} = -1$ models. Note that $\mathcal{P}_2$ is now very poorly constrained and also that $\sigma_8$ increases with the inclusion of lensing data unlike before.} \label{fig:contour3}
\end{figure*}

In this section we provide the 2D posterior distribution marginalised cosmological contour plots between $\mathcal{P}_1$, $\mathcal{P}_2$, and $\sigma_8$ in \autoref{fig:contour1}, as well as $A_{\rm lens}$ in \autoref{fig:contour3}, and also for $\mathcal{P}_3$, $\mathcal{P}_4$, $\sigma_8$, and $w_{\rm de}$ in \autoref{fig:contour2}.
\twocolumngrid
\clearpage
\bibliographystyle{apsrev4-1}
%\bibliography{bibfile.bib}
\bibliography{draft.bbl}
\end{document}